\documentclass[12pt]{article}
\usepackage{amsmath,amssymb,epsfig}
\makeatletter \@addtoreset{equation}{section} \makeatother
\renewcommand{\theequation}{\thesection.\arabic{equation}}
\newcommand{\ba}{\begin{array}}
\newcommand{\ea}{\end{array}}
\newcommand{\beq}{\begin{equation}}
\newcommand{\eeq}{\end{equation}}
\newcommand{\bea}{\begin{eqnarray}}
\newcommand{\eea}{\end{eqnarray}}

\def\bce{\begin{center}}
\def\ece{\end{center}}

\def\nonu{\nonumber}

\def\be{\beta}

\newcommand{\tr}{\mbox{Tr}}

\def\eps6{{\displaystyle \mathop{\epsilon}^{6}}{}}

\def\nab6{{\displaystyle \mathop{\nabla}^{6}}{}}

\def\0{{\sst{(0)}}}
\def\1{{\sst{(1)}}}
\def\2{{\sst{(2)}}}
\def\3{{\sst{(3)}}}
\def\4{{\sst{(4)}}}
\def\5{{\sst{(5)}}}
\def\6{{\sst{(6)}}}
\def\7{{\sst{(7)}}}
\def\8{{\sst{(8)}}}

\def\ba{\begin{array}}
\def\ea{\end{array}}
\def\beq{\begin{equation}}
\def\eeq{\end{equation}}
\def\be{\begin{equation}}
\def\ee{\end{equation}}

\def\tr{\mathop{\rm tr}}

\def\eps{\epsilon}

\def\ba{\begin{array}}
\def\ea{\end{array}}
\def\beq{\begin{equation}}
\def\eeq{\end{equation}}
\def\be{\begin{equation}}
\def\ee{\end{equation}}

\def\tr{\mathop{\rm tr}}

\def\eps{\epsilon}

\newcommand{\bean}{\begin{eqnarray*}}
\newcommand{\eean}{\end{eqnarray*}}

\begin{document}
\thispagestyle{empty} \addtocounter{page}{-1}
\begin{flushright}
KIAS-P07034 \\
\end{flushright}

\vspace*{1.3cm}

\centerline{ \Large \bf  Meta-Stable Brane Configurations with
Seven NS5-Branes}
\vspace*{1.5cm}
\centerline{{\bf Changhyun Ahn} 
} 
\vspace*{1.0cm} 
\centerline{\it 
Department of Physics, Kyungpook National University, Taegu
702-701, Korea} 
\vspace*{0.8cm} 
\centerline{\tt ahn@knu.ac.kr} 
\vskip2cm

\centerline{\bf Abstract}
\vspace*{0.5cm}

We present the intersecting brane configurations 
consisting of NS-branes, 
D4-branes(and anti D4-branes) and O6-plane,
of type IIA string
theory corresponding to 
the meta-stable nonsupersymmetric vacua in four dimensional ${\cal N}=1$
supersymmetric $SU(N_c) \times SU(N_c') \times SU(N_c'')$ gauge theory with   
a symmetric tensor field, a conjugate symmetric tensor field  and 
bifundamental fields.   
We also describe the intersecting brane configurations 
of type IIA string
theory corresponding to 
the nonsupersymmetric meta-stable vacua in the above gauge theory with   
an antisymmetric tensor field, a conjugate symmetric tensor field, 
eight fundamental flavors and bifundamentals. 
These brane configurations consist of 
NS-branes, 
D4-branes(and anti D4-branes),
D6-branes and O6-planes.

\baselineskip=18pt
\newpage
\renewcommand{\theequation}
{\arabic{section}\mbox{.}\arabic{equation}}

\section{Introduction}

The dynamical supersymmetry breaking in meta-stable vacua \cite{ISS} 
arises 
in the ${\cal N}=1$ SQCD with massive fundamental flavors 
where the masses for quarks 
are much smaller than the dynamical scale of gauge
sector. See also the review paper  \cite{IS} for recent developments
on supersymmetry breaking.
In the magnetic dual theory with massless quarks, 
the superpotential has a cubic
interaction term between the meson and quarks. For the corresponding 
brane configuration where there exist two NS-branes, D4-branes and
D6-branes, 
see the reference \cite{GK98}. 
Due to the extra mass term
for quarks in the magnetic superpotential with massive quarks, 
not all F-term equations can be satisfied. Then the
supersymmetry is broken by rank condition \cite{ISS}.
Therefore, both the mass term for the quarks in the superpotential 
and the magnetic dual
procedure are crucial in order 
to construct the new meta-stable supersymmetry
breaking vacua.  
The meta-stable brane configurations of type IIA string theory
corresponding to ${\cal N}=1$ SQCD with massive fundamental flavors 
have been found in \cite{OO,FGU,BGHSS}. See also \cite{Ahn06} which
has dealt with the case where there exists an extra adjoint matter.
 
To construct brane configurations that we will describe, it is very
useful to decompose $9 + 1$ dimensional space time 
as 
\bea
(x^0, x^1, x^2, x^3), \qquad (x^4, x^5),  \qquad
x^6, 
\qquad x^7, \qquad (x^8, x^9).
\nonu
\eea
Then $3+1$ dimensions labeled by $(x^0, x^1, x^2, x^3)$ are common all
the branes and we introduce two complex planes 
\bea
 v = x^4 + i x^5, \qquad w = x^8 + i x^9.
\nonu
\eea 
There are two types of NS-branes which stretch in $(x^0, x^1, x^2, x^3)$
as well as following directions
\bea
NS5-\mbox{brane} & : & v, 
\nonu \\
NS5'-\mbox{brane} & : & w. 
\nonu
\eea
These branes preserve ${\cal N}=2$ supersymmetry in $3+1$ dimensions.
The NS-brane configuration of interest is given by 
a single NS5-brane extended $v$ and localized at $x^6=0$ and two
NS5'-branes at $(v=v_1, x^6= y_1)$ and $(v=v_2, x^6= y_2)$ respectively.
Now we add $N_c'$ D4-branes stretched between the NS5-brane and 
one of the NS5'-branes and $N_c$ $\overline{D4}$-branes stretched
between the NS5-brane and the other NS5'-brane in order to break 
the supersymmetry.
These D4-branes and  $\overline{D4}$-branes stretch in $(x^0, x^1,
x^2, x^3)$ and have finite interval along $x^6$-direction.
Then this brane construction is drawn in Figure 1A found in \cite{GK}.
In the Figures of this paper, 
the vertical axis is $v$-direction, the horizontal axis is 
$x^6$-direction and the other orthogonal direction coming out of the
page is $w$-direction.

\begin{figure}[ht]
   \epsfxsize=3.5in 
\centerline{\epsffile{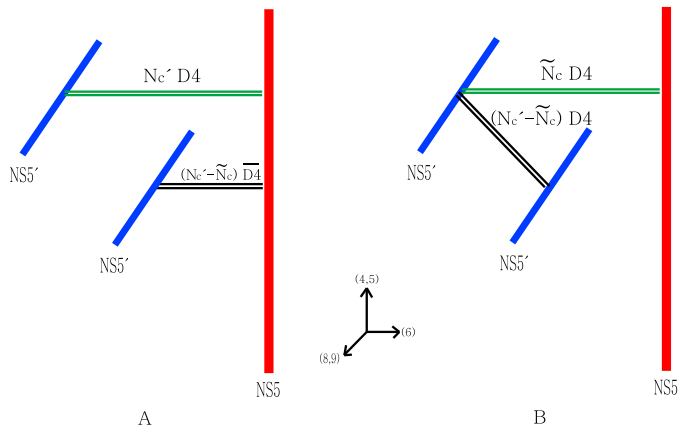}}
   \caption[FIG. \arabic{figure}.]{ 
The  
 ${\cal N}=1$ magnetic brane configuration for the gauge group 
$SU(\widetilde{N}_c=N_c'-N_c) \times SU(N_c')$ with D4-
and $\overline{D4}$-branes(1A) and with 
a misalignment between D4-branes(1B) when the NS5'-branes of 1A are close to
each other.  }
\end{figure}

Recently, Giveon and Kutasov \cite{GK} have found 
the type IIA brane configuration consisting of 
two NS5'-branes, a single NS5-brane, 
D4-branes and anti-D4-branes($\overline{D4}$-branes).
The gravitational interaction for the D4-branes 
in the background of the NS5-branes has led to 
the phase structures in different regions of the 
parameter space. The meta-stable vacua of \cite{ISS} occur in some
region of parameter space when the D4-branes and
$\overline{D4}$-branes can decay(by taking the distance between the
NS5'-branes very small)  
and the geometric misalignment of
flavor D4-branes arises and this feature is shown in Figure 1B. 

One advantage of this construction is the
fact that there are no D6-branes and this enables us to write the
magnetic superpotential in simple form. One disadvantage of 
this construction is that one has 
to increase the number of NS-branes, compared with  the
description of \cite{OO,FGU,BGHSS}, since the
role of D6-branes is replaced by the role of NS5'-brane.  
The mass term in the magnetic superpotential in this approach
corresponds to the relative displacement of two NS5'-branes 
along the 45 directions and the dual quarks can be represented by the 
bifundamentals of product gauge group since there are three
NS-branes. 
When the NS5'-brane is replaced by coincident D6-branes, then the
magnetic meta-stable 
brane configuration will lead to the one in \cite{OO,FGU,BGHSS}. 

Adding an orientifold 4-plane only to the brane configuration 
of \cite{GK} implies that the gauge group is a product of a symplectic
group and an orthogonal group and the geometric misalignment of flavor 
D4-branes \cite{Ahn07-5} together with a replacement of NS5'-branes by coincident
D6-branes leads to the brane configuration of \cite{Ahn06-1}. 
In the standard supersymmetric electric brane configuration of
type IIA string theory, 
the ${\cal N}=1$ product gauge group theory \cite{ILS,BIWW} 
is realized by three
NS-branes, two kinds of D4-branes, and two kinds of 
D6-branes \cite{BH}.  
When an orientifold 6-plane is added into the brane configuration of 
\cite{GK}, the gauge group is a product of two unitary groups 
with extra matters as well as bifundamentals. In this case, 
the type IIA brane configuration consists of 
five(the number of NS-branes plus the number 
of kinds of D6-branes in
\cite{BH}) NS-branes, 
D4-branes, $\overline{D4}$-branes and O6-plane.
Similarly,  the geometric misalignment of flavor 
D4-branes \cite{Ahn07-5} with a replacement of NS5'-branes by coincident
D6-branes leads to the brane configurations of \cite{Ahn07} or 
\cite{Ahn07-1} depending on the O6-plane charge. 

One can generalize the work of \cite{GK} further by adding more NS-branes to
the brane configuration of \cite{GK}.
For the ${\cal N}=1$ triple product group gauge theory, the
supersymmetric electric brane configuration in type IIA string theory 
consists of  four
NS-branes, three kinds of D4-branes, and three kinds of 
D6-branes \cite{BH,AT97} in the conventional brane configuration.
Then the triple product gauge group theory with bifundamentals only
can be realized by four NS-branes, three kinds of 
D4-branes and $\overline{D4}$-branes \cite{Ahn07-6} since there are no
quarks.
The meta-stable vacua of \cite{Ahn07-3} occur in some
region of parameter space when the D4-branes and
$\overline{D4}$-branes can decay and the geometric misalignment of
flavor D4-branes arises.

Adding an orientifold 4-plane only to the brane configuration 
of \cite{Ahn07-6} leads to the fact that the gauge group is a triple 
product of a symplectic
group and an orthogonal group alternatively 
and the geometric misalignment of flavor 
D4-branes \cite{Ahn07-6} together with a replacement of NS5'-branes by coincident
D6-branes reduces to the brane configuration of \cite{Ahn07-2}. 
One can add an orientifold 6-plane into the brane configuration of 
\cite{BH,AT97} together with extra outer NS-branes in order to keep
the number of gauge group factor. 
Then
the type IIA brane configuration consists of 
six NS-branes, 
D4-branes, $\overline{D4}$-branes and O6-plane.
The geometric misalignment of flavor 
D4-branes \cite{Ahn07-6} with a replacement of NS5'-branes by coincident
D6-branes leads to the brane configurations of \cite{Ahn07-3}. 
Different O6-plane charge will give rise to other brane configuration 
of \cite{Ahn07-3} by a replacement of NS5'-branes by D6-branes. 

In this paper, we  present the intersecting 
brane configurations of type IIA string
theory corresponding to 
the new meta-stable nonsupersymmetric vacua in four dimensional ${\cal
N}=1$ triple product  gauge theory with matters.
For the ${\cal N}=1$ product gauge group theory with extra matters as well
as bifundamentals, the
supersymmetric electric brane configuration in type IIA string theory 
consists of  five
NS-branes, two kinds of D4-branes, and two kinds of 
D6-branes as well as O6-plane \cite{Ahn07-4}.

When the two outer  NS5-branes are added into the brane configuration of 
\cite{Ahn07-4} and removing the D6-branes completely, 
the gauge group is a product of three unitary groups 
with extra matters as well as bifundamentals. In this case, 
the type IIA brane configuration consists of 
seven NS-branes(i.e., the number of NS-branes plus the number 
of kinds of D6-branes in \cite{Ahn07-4}), 
D4-branes, $\overline{D4}$-branes and O6-plane.
For a given supersymmetric electric gauge theory which does not have
any superpotential, 
one takes both the mass deformation for bifundamentals and its 
Seiberg dual. Then we construct the meta-stable 
brane configurations in type IIA string theory.    
Furthermore, the same gauge theory with more complicated matters
is realized by 
seven NS-branes, 
D4-branes, $\overline{D4}$-branes, D6-branes and two kinds of 
O6-planes. 

In section 2, we describe the type IIA brane configuration 
corresponding
to the electric theory based on the ${\cal N}=1$ $SU(N_c) \times
SU(N_c') \times SU(N_c'')$ 
gauge theory 
with   
a symmetric tensor field, a conjugate symmetric tensor field  and 
bifundamental fields
and deform this theory by adding the mass term
for the bifundamental. 
We construct the three different 
dual magnetic theories by taking the Seiberg
dual for each gauge group factor. 
Then we consider the nonsupersymmetric meta-stable
minimum  and present 
the corresponding intersecting brane configurations of type IIA string
theory.

In section 3, we discuss the type IIA brane configuration
corresponding
to the electric theory based on the ${\cal N}=1$ $SU(N_c) \times
SU(N_c') \times SU(N_c'')$ 
gauge theory 
with 
an antisymmetric tensor field, a conjugate symmetric tensor field, 
eight fundamental flavors and bifundamentals
and deform this theory by adding the mass term
for the bifundamental. 
Then we construct the  corresponding 
dual magnetic theories  by taking the Seiberg
dual for each gauge group factor. 
We consider the nonsupersymmetric meta-stable
minimum  and present 
the corresponding intersecting brane configurations of type IIA string
theory. 

In section 4, we summarize what we have found in this paper and 
make some comments for the future directions
\footnote{A replacement of D6-branes with NS5'-brane corresponds to
  the gauging of the flavor group
of the gauge theory realized 
on the D4-branes
and this replacement 
might be useful to construct the phenomenological model
building.}.  

\section{Meta-stable brane configurations-I}

The type IIA brane configuration  corresponding to 
${\cal N}=1$ supersymmetric gauge theory with
gauge group
\bea
SU(N_c) \times SU(N_c') \times SU(N_c'') 
\label{gag}
\eea
and with a symmetric tensor field $S$ charged under $({\bf \frac{1}{2}
N_c(N_c+1), 1, 1})$, 
a field $F$ charged under
$({\bf N_c, \overline{N_c'}, 1})$, a field $G$ charged under
$({\bf 1, N_c', \overline{N_c''}})$, and their conjugates 
$\widetilde{S}, \widetilde{F}$ and $\widetilde{G}$ 
can be described by 
the left $NS5_L'$-brane(012389), 
the  
NS5-brane(012345), the right $NS5_R'$-brane(012389)(and their
mirrors), the middle NS5-brane(012345),
$N_c$-, $N_c'$-  and $N_c''$-color D4-branes(01236)(and their mirrors) 
as well as
the $O6^{+}$-plane(0123789). 
The $O6^{+}$-plane acts as $(x^4,x^5,x^6) \rightarrow
(-x^4,-x^5,-x^6)$ and has RR charge $+4$
\footnote{When we say about NS-branes(i.e., NS5-brane or
  NS5'-brane) in this section, 
those NS-branes are located at the region of $x^6 >0$. 
We assume that their mirrors behave according to this ${\bf Z}_2$ symmetry
of $O6^{+}$-plane. 
That is, there exist three NS-branes:
$NS5_L'$-brane,
NS5-brane and $NS5_R'$-brane from Figure 2A.}.
See also the relevant works \cite{LLL,LL} which have dealt with the
gauge theory realized by the first factor in (\ref{gag}) with a
symmetric tensor, a conjugate symmetric tensor and fundamentals. 

Let us place an $O6^{+}$-plane at the origin $x^6=0$
and the $x^6$ 
coordinates for the $NS5_L'$-brane, the NS5-brane and the $NS5_R'$-brane 
are given by $x^6=y_1, y_1+y_2, y_1+y_2+y_3$
respectively. The locations for their mirrors can be understood similarly.
The $N_c$ D4-branes 
are suspended between the 
$NS5_L'$-brane and its mirror, 
the $N_c'$ D4-branes 
are suspending between the 
$NS5_L'$-brane and the NS5-brane(and their mirrors), and 
the $N_c''$ D4-branes  
are suspended between the NS5-brane and the $NS5_R'$-brane(and their mirrors).
The fields $S$ and $\widetilde{S}$  correspond to 4-4 strings connecting 
the $N_c$-color D4-branes with $x^6 < 0$ with $N_c$-color D4-branes
with $x^6 > 0$.
The fields $F$ and $\widetilde{F}$  correspond to 4-4 strings connecting 
the $N_c$-color D4-branes with $N_c'$-color D4-branes while
the fields $G$ and $\widetilde{G}$  correspond to 4-4 strings connecting 
the $N_c'$-color D4-branes with $N_c''$-color D4-branes.
We draw this ${\cal N}=1$ supersymmetric 
electric brane configuration in Figure 2A for the vanishing mass
for the fields $G$ and $\widetilde{G}$. 

\begin{figure}[ht]
   \epsfxsize=5.0in 
\centerline{\epsffile{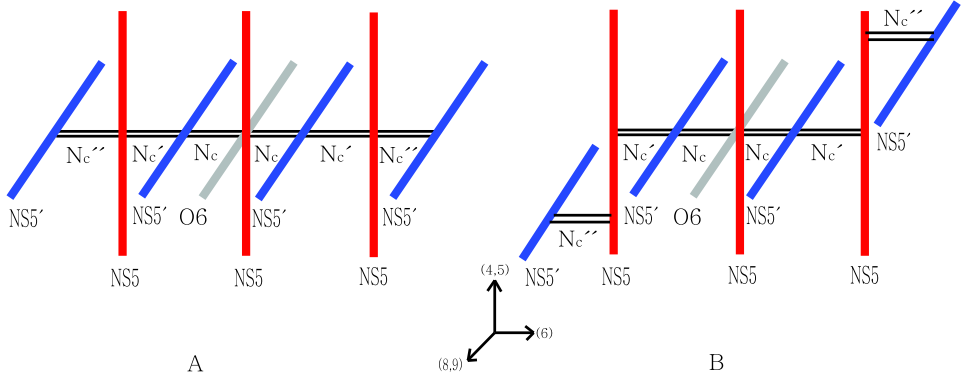}}
   \caption[FIG. \arabic{figure}.]{ 
The  ${\cal N}=1$ supersymmetric 
electric brane configuration for the gauge group $SU(N_c) \times
SU(N_c') \times SU(N_c'')$ and  a symmetric tensor $S$, a conjugate
symmetric tensor $\widetilde{S}$, 
the bifundamentals $F, \widetilde{F}, G$
and $\widetilde{G}$  with vanishing(2A) and 
nonvanishing(2B) mass
for the fields $G$ and $\widetilde{G}$. 
The $NS5_R'$-brane together 
with $N_c''$ D4-branes is moving to $+v \equiv x^4 + i x^5$ 
direction in 2B(and their 
mirrors to $-v$ direction). }
\end{figure}

The gauge couplings of $SU(N_c)$, $SU(N_c')$ and $ SU(N_c'')$
are given by a string coupling constant $g_s$, a string scale $\ell_s$ 
and the $x^6$ coordinates $y_i$ for three NS-branes through
\bea
g_1^2 =\frac{g_s \ell_s}{y_1}, \qquad 
g_2^2 = \frac{g_s \ell_s}{y_2}, \qquad
g_3^2=\frac{g_s \ell_s}{y_3}.
\nonu
\eea
As $y_3$ goes to $\infty$, the  
$SU(N_c'')$ gauge group becomes a
global symmetry and the theory leads to the super QCD with the gauge group
$SU(N_c) \times SU(N_c')$ and $N_c''$ flavors 
in the fundamental representation of second gauge group factor \cite{Ahn07-4}.

Now we describe three different magnetic dual theories by taking each
corresponding mass deformation.

\subsection{${\cal N}=1$ 
$SU(N_c) \times SU(\widetilde{N}_c') \times SU(N_c'')$ magnetic theory}

Let us  take 
the mass deformation by moving the $N_c''$ D4-branes with
$NS5_R'$-brane 
to $+v$ direction
in the electric theory and 
then consider the Seiberg dual for the second gauge group in this
subsection \footnote{When we move NS-branes or D4-branes,
we only refer to the original NS-branes or D4-branes located at the
positive $x^6$ region. The movement of
their mirrors is evident if we remember the action of $O6^{+}$-plane we
mentioned before. Then, although we describe only the movement of 
original NS-branes
or D4-branes, their mirrors also are moving  automatically, 
according to the action of
$O6^{+}$-plane.}.

$\bullet$ Mass deformation by $G$ and $\widetilde{G}$

There is no electric superpotential in Figure 2A.
It is known in \cite{Ahn07-4} that the classical superpotential 
without $NS5_R'$-brane(and its mirror) for the particular orientation
for NS-branes in Figure 2A is vanishing.
Now we are adding the $NS5_R'$-brane(and its mirrors) 
into the brane configuration of 
\cite{Ahn07-4} together with $N_c''$ D4-branes(and their mirrors).
Since the angle between the NS5-brane and the extra $NS5_R'$-brane
is $\frac{\pi}{2}$ in Figure 2A, the mass for the extra adjoint field $\mu''$
corresponding to the
gauge group $SU(N_c'')$
becomes large and after integrating out this adjoint field, then
the extra classical
superpotential term which has a factor $\frac{1}{\mu''}$ will vanish.  

Now let us deform this theory.
Displacing the two NS5'-branes relative each other in the $+v$
direction corresponds to turning on a quadratic
mass-deformed electric superpotential
for the field $G$ as follows:
\bea
W_{def} = m G \widetilde{G} (\equiv m \tr \Phi''), \qquad m = 
\frac{\Delta x}{\ell_s^2}
\label{Mass}
\eea
where 
the second gauge group indices in $G$ and $\widetilde{G}$ 
are contracted, the third gauge group indices in them are free 
and the mass $m$ is related to the relative 
displacement $\Delta x$ between the two NS5'-branes where $x \equiv x^5$.
The gauge-singlet $\Phi''$, which has two free indices on the third gauge
group
$SU(N_c'')$, 
is in the 
adjoint representation for the third  gauge group $SU(N_c'')$, 
i.e., ${(\bf 1, 1, {N_c''}^2-1)  \oplus (1,1,1)}$ 
under the gauge group. The trace in (\ref{Mass}) 
is taken over the third gauge group indices and 
the gauge singlet $\Phi''$ is a $N_c'' \times N_c''$ matrix.
The $NS5_R'$-brane together with $N_c''$-color D4-branes 
is moving to the $+v$ direction  for
fixed other branes during this particular 
mass deformation(and their mirrors to
$-v$ direction). 
Then the $x^5$ coordinate 
of $NS5_L'$-brane is equal to
zero
and the $x^5$ coordinate of $NS5_R'$-branes is given by 
$ \Delta x$.
Giving an expectation value to the meson field $\Phi''$
corresponds to recombination of $N_c'$- and $N_c''$- color 
D4-branes, which will become $N_c''$- or $N_c'$-color D4-branes
in Figure 2A such that they are suspended between 
the $NS5_L'$-brane and the $NS5_R'$-brane 
and moving them into the $w \equiv x^8 + i x^9$
direction. We assume 
that the number of colors satisfies
\bea
N_c+N_c'' \geq N_c' \geq N_c.
\label{color}
\eea

Now 
we draw this ${\cal N}=1$ supersymmetric electric 
brane configuration in Figure 2B for nonvanishing mass
for the fields $G$ and $\widetilde{G}$. 
The geometric configuration for three NS-branes in Figure 2B 
is exactly the same as the first
three NS-branes in Figure 1B of \cite{Ahn07-6}.

$\bullet$ Seiberg dual

Now it is ready to take the magnetic gauge theory from 
the above specific deformed electric gauge theory.
This is a necessary step for the construction of 
meta-stable brane configuration.
By applying the Seiberg dual to the second gauge group $SU(N_c')$ factor in 
(\ref{gag}), the $NS5_{L,R}'$-branes can be placed at the
outside of the three NS5-branes, as in Figure 3.
Starting from Figure 2B and 
interchanging the $NS5_L'$-brane and the NS5-brane(and their mirrors),
one obtains the Figure 3A with D4- and $\overline{D4}$-branes.

\begin{figure}[ht]
   \epsfxsize=5.0in 
\centerline{\epsffile{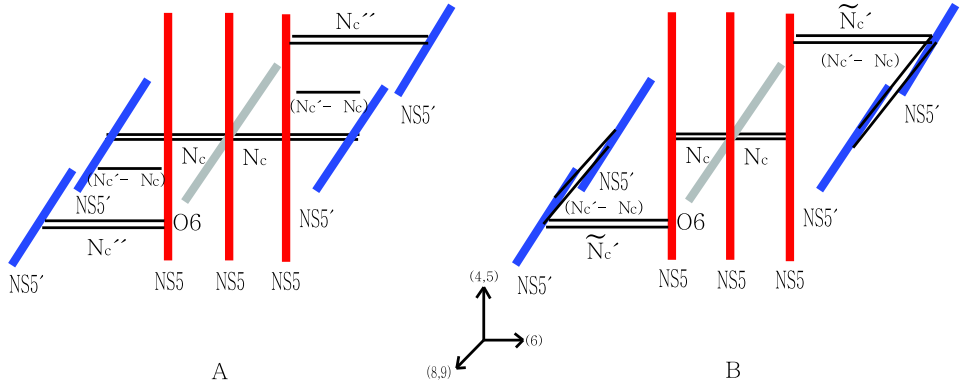}}
   \caption[FIG. \arabic{figure}.]{ 
The  ${\cal N}=1$ magnetic brane configuration for the gauge group 
$SU(N_c) 
\times SU(\widetilde{N}_c'=N_c+N_c''-N_c') \times SU(N_c'')$ 
corresponding to Figure 2B with D4-
and $\overline{D4}$-branes(3A) and with 
a misalignment between D4-branes(3B) when the NS5'-branes of 3A are close to
each other. The number of tilted D4-branes in 3B can be written as
$N_c'-N_c=N_c''-\widetilde{N}_c'$. The notation for the anti-D4-branes
is used with the bar on the number of those branes in Figure 3A.}
\end{figure}

Before arriving at the Figure 3A, there exists an intermediate 
step where 
the $(N_c''-N_c'+N_c)$ D4-branes are connecting between the 
NS5-brane and the  $NS5_L'$-brane,  
$N_c''$ D4-branes are connecting between the  $NS5_L'$-brane and   
$NS5_R'$-brane(and their mirrors) as well as $N_c$ D4-branes between
the
NS5-brane and its mirror.
By reconnecting the $N_c''$ D4-branes 
connecting between the 
NS5-brane and the  $NS5_L'$-brane
with  
the $N_c''$ D4-branes connecting between  
$NS5_L'$-brane 
and the $NS5_R'$-brane and moving those combined
$N_c''$ 
D4-branes
to $+v$-direction(and their mirrors to $-v$ direction), 
one gets the final Figure 3A where we are left with 
$(N_c'-N_c)$ 
anti-D4-branes between the NS5-brane and   
$NS5_L'$-brane.

When two NS5'-branes in Figure 3A are close to each other, one arrives at 
the Figure 3B
by realizing that the number of $N_c''$
D4-branes connecting between NS5-brane and $NS5_R'$-brane in Figure
3A can
be rewritten as $(N_c'-N_c)$ plus $\widetilde{N}_c'$. 
The brane configuration consisting of NS5-brane and two NS5'-branes in
Figure 3B is
exactly the same as the first three NS-branes 
in Figure 3B of \cite{Ahn07-6}.  
Moreover if we change the $O6^{+}$-plane into the $O6^{-}$-plane,
replace the number of color $N_c$ by $2N_c$, and  remove the middle
NS5-brane 
in Figure 3, then this is exactly the same as the Figure 12 of \cite{Ahn07-6}.

The dual gauge group from the computation of linking number counting 
during the brane motion is given by 
\bea
SU(N_c) \times SU(\widetilde{N}_c'=N_c+N_c''-N_c') \times SU(N_c''). 
\label{dualnew}
\eea
The matter contents in the magnetic theory are  a symmetric tensor field $S$ 
charged under $({\bf \frac{1}{2}
N_c(N_c+1), 1, 1})$, the field $f$ 
 charged under
$({\bf N_c, \overline{\widetilde{N}_c'}, 1})$, a field $g$ charged under
$({\bf 1, \widetilde{N}_c', \overline{N_c''}})$, and their conjugates 
$\widetilde{S}, \widetilde{f}$ and $\widetilde{g}$ under the dual gauge group
(\ref{dualnew})
and  
the gauge-singlet $\Phi''$ for the second dual gauge group in the 
adjoint representation for the third dual gauge group, 
i.e.,  ${(\bf 1, 1, {N_c''}^2-1)  \oplus (1,1,1)}$ under the 
dual gauge group.
Then the  gauge singlet $\Phi''$ which was defined in (\ref{Mass}) 
is nothing but a $N_c'' \times N_c''$ matrix.

$\bullet$ Meta-stable brane configuration

The combination of cubic superpotential in the magnetic theory 
with the deformed mass term (\ref{Mass}) 
is given by
\footnote{In general, there are also 
the extra terms in the superpotential
$\Phi_1 f g + \Phi f \widetilde{f} + \Phi_3 \widetilde{f} 
\widetilde{g}$ where we define 
$\Phi_1 \equiv F G$,  
$\Phi \equiv F \widetilde{F}$ and $\Phi_3 \equiv \widetilde{F} \widetilde{G}$, 
coming from 
different bifundamentals. However, the F- term conditions,
$\Phi_1 g + \Phi \widetilde{f}=0=\Phi
f + \Phi_{3} \widetilde{g}$ lead to 
$<\Phi_1>=<\Phi_3>=<f>=<\widetilde{f}>=0$. 
Therefore, these extra terms do not
contribute to the one loop computation up to quadratic order.}
\bea
W_{dual} = \Phi'' g \widetilde{g} + m \tr \Phi''. 
\label{superpo1new}
\eea
Here the magnetic fields $g$ and $\widetilde{g}$  
correspond to 4-4 strings connecting 
the $\widetilde{N}_c'$-color D4-branes that are 
connecting between the NS5-brane
and the $NS5_R'$-brane in Figure 3B with $N_c''$-flavor 
D4-branes which  are realized 
as corresponding D4-branes in Figure 3A.
Although the $N_c''$ D4-branes in Figure 3A cannot move any
directions,
the tilted $(N_c'-N_c)$-flavor D4-branes can move 
$w$ direction in Figure 3B.
The remaining upper $\widetilde{N}_c'$ D4-branes are fixed also and cannot 
move any directions. 
It is useful for the transition from Figure 3A to Figure 3B to note that 
there is a decomposition 
\bea
N_c''=(N_c'-N_c)+\widetilde{N}_c'.
\nonu
\eea
The $N_c$ D4-branes between the NS5-brane and the middle NS5-brane can
slide along the $v$ direction(and their mirrors to the opposite direction).

The brane configuration for zero mass for the bifundamental $G$ and $\widetilde{G}$,
which has only a cubic superpotential in (\ref{superpo1new}),
can be seen from the Figure 3A by moving
the upper  NS5'-brane(or $NS5_R'$-brane) 
together with $N_c''$ color D4-branes 
into the origin $v=0$(and their mirrors): massless limit 
after the Seiberg dual with massive case.
Then the number of dual colors for D4-branes 
becomes $N_c$ between the NS5-brane and its mirror, 
$\widetilde{N}_c'$ between the NS5-brane and the $NS5_L'$-brane
and $N_c''$ 
between the $NS5_L'$-brane and the $NS5_R'$-brane.
Or starting from the Figure 2A and moving the 
NS5-brane to the left all the
way past the $NS5_L'$-brane(and their mirrors),
one also obtains the corresponding magnetic brane configuration
for massless case: the Seiberg dual with massless case.

The brane configuration in Figure 3A is stable as long as the
distance $\Delta x$ between the upper NS5'-brane and 
the lower NS5'-brane is large.
The fundamental strings are stretched between D4 and
$\overline{D4}$-branes in Figure 3A
and the lowest exciting states 
are given by open string tachyon. Now the effective mass for 
open string tachyon can be computed via near horizon geometry 
of the NS5-branes and is given by the result (3.16) of \cite{GK}.
Then for $\Delta x > \pi l_s$, the open string tachyon 
is massive everywhere and the Figure 3A is locally stable and global 
ground state of the system in this regime at least classically.
 
If they are close to each other, then this brane
configuration of Figure 3A is unstable to decay and leads to 
the meta-stable brane configuration in Figure
3B.
One regards these two type IIA brane configurations, in the zero limit
of $g_s$, as 
particular states in the
magnetic gauge theory, when $\Delta x$ or $m$ is very small,  
with the gauge group (\ref{dualnew}) and
superpotential (\ref{superpo1new}), along the line of \cite{GK}.

One can perform similar analysis done in \cite{GK,Ahn07-5,Ahn07-6} 
for our brane configuration 
since one can take into account the behavior of
parameters geometrically in the presence of $O6^{+}$-plane.
Then the  upper tilted $(N_c''-\widetilde{N}_c')$ flavor D4-branes of 
straight brane configuration
of
Figure 3B can bend due to the fact that there exists an attractive
gravitational interaction
between those flavor D4-branes and the NS5-brane from the DBI action
as long as $y_1$, the $x^6$ coordinate of NS5-brane, is very large.
Then the mirror of NS5-brane and the middle NS5-brane do 
not affect those flavor D4-branes.
On the other hand, 
if $y_1$ goes to zero, then the mirror of NS5-brane plus the middle
NS5-brane
play the role
of enhancing the strength for the  
NS5-branes and will affect both the energy of bending curve, 
$E_{curved}$,  and $\Delta x$ as in \cite{GK}. 
Of course, their mirrors, the lower tilted
$(N_c''-\widetilde{N}_c')$ flavor D4-branes of 
straight brane configuration
of
Figure 3B can bend and their trajectories connecting 
two NS5'-branes should be preserved under the $O6^{+}$-plane action.

$\bullet$ Gauge theory at small $\Delta x$ 

The low energy dynamics of the magnetic brane configuration 
can be described by the ${\cal N}=1$ supersymmetric gauge theory
with gauge group (\ref{dualnew})
and the gauge couplings for the three gauge group factors are
given by
\bea
g_{1,mag}^2  = \frac{g_s \ell_s}{(y_1+y_2)}, \qquad 
g_{2,mag}^2 = \frac{g_s \ell_s}{y_2}, \qquad
g_{3,mag}^2  = \frac{g_s \ell_s}{(y_3+y_2)}
\nonu
\eea
where
$y_1$ is the $x^6$ coordinate of $NS5_L'$-brane,
$y_1+y_2$ is the $x^6$ coordinate of NS5-brane, and
$y_1+y_2+y_3$ is the $x^6$ coordinate of $NS5_R'$-brane in Figure 3,
as we described before.
The dual gauge theory has  a meson $\Phi''$  and 
bifundamentals $f, \widetilde{f}, g$ and $\widetilde{g}$ as well as 
a symmetric tensor $S$ and its conjugate $\widetilde{S}$ 
under the dual gauge
group (\ref{dualnew}) and the superpotential 
corresponding to the Figures 3A and 3B is given by 
\bea
W_{dual} = h \Phi'' g \widetilde{g} - h \mu^2 \tr \Phi'', \qquad h^2 = g_{3,
  mag}^2,
\qquad \mu^2 = -\frac{\Delta x}{ 2\pi g_s \ell_s^3}.
\label{wdual}
\eea
Then the product 
$ g \widetilde{g}$ is a $\widetilde{N}_c' \times \widetilde{N}_c'$ 
matrix where the third gauge group indices for $g$ and $\widetilde{g}$ 
are contracted with those
of $\Phi''$ and $\Phi''$ is a 
$N_c'' \times N_c''$ matrix.
The product $g \widetilde{g}$ has the same representation for the 
product of quarks
and 
the third gauge group indices for the field $\Phi''$ play the
role of the flavor indices.

When the upper NS5'-brane(or $NS5_R'$-brane) 
is replaced by coincident $N_c''$ 
D6-branes in Figure 3B, this brane configuration looks similar to the
Figure 5 
found in \cite{Ahn07-4} where the gauge group was given by 
$SU(n_c) \times SU(n_f'+n_c-n_c')$ 
with $n_f'$ multiplets, bifundametals, a symmetric
flavor,
a conjugate symmetric flavor and various gauge singlets. 
Then the present $N_c$ for the first gauge group factor 
corresponds to the $n_c$,
the $N_c'$ for the second gauge group factor corresponds to $n_c'$,
and finally
the $N_c''$ corresponds to the $n_f'$ of \cite{Ahn07-4}. 
The location of $NS5_R'$-brane of Figure 3B 
is in the right hand side of NS5-brane
while the location of $n_f'$ D6-branes of Figure 5 in \cite{Ahn07-4}
is in the left hand side of NS5-brane.

When $y_1$ goes to zero, this meta-stable brane configuration of
Figure 3B where there are three NS5-branes at the origin 
reduces to  the Figure 4 of \cite{Ahn07-5} where there is a single
NS5-brane at the origin.

Therefore, the F-term equation, the derivative of $W_{dual}$ in
(\ref{wdual}) 
with respect to the
meson field $\Phi''$ cannot be satisfied if the $N_c''$ exceeds
$\widetilde{N}_c'$.
So the supersymmetry is broken.   
That is, 
there exist three equations from F-term conditions:
$
g^a \widetilde{g}_b -\mu^2 \delta^a_b =0$ and $ \Phi'' g =0=\widetilde{g} \Phi''$.
Then the solutions for these, as usual,
are given by 
\bea
<g>   = 
\left(
\begin{array}{c}
\mu  {\bf 1}_{\widetilde{N}_c'}  \\
0
\end{array}
\right), 
\qquad
<\widetilde{g}>   = 
\left(
\begin{array}{cc}
\mu  {\bf 1}_{\widetilde{N}_c'} & 0  \\
\end{array}
\right), 
\qquad
<\Phi''> =
 \left(
\begin{array}{cc}
0  & 0  \\
0 & \Phi_0''  {\bf 1}_{(N_c''-\widetilde{N}_c')} 
\end{array}
\right) 
\label{point20}
\eea
where the zero of $<g>$ is a $
(N_c''-\widetilde{N}_c') \times \widetilde{N}_c'$ 
matrix, the zero of $<\widetilde{g}>$ is a
$\widetilde{N}_c' \times (N_c''-\widetilde{N}_c') $ matrix and 
the zeros of $<\Phi''>$ are $\widetilde{N}_c' \times \widetilde{N}_c'$,
$\widetilde{N}_c' \times 
(N_c''-\widetilde{N}_c')$ and $(N_c''-\widetilde{N}_c') \times
\widetilde{N}_c'$ 
matrices.
Then one can expand these fields around on a point (\ref{point20}), as
in \cite{ISS,Shih} and one arrives at the relevant superpotential
up to quadratic order in the fluctuation. 
At one loop, the effective potential $V_{eff}^{(1)}$ for $\Phi_0''$
leads to the positive value for $m_{\Phi_0''}^2$ implying that these
vacua are stable.
The gauge theory analysis where the theory will be strongly coupled in
the IR region 
is only valid in the regime where 
$\Delta x$ is smaller than $\exp(-\frac{C}{g_s})$ with some positive
constant $C$. 

\subsection{${\cal N}=1$ 
$SU(N_c) \times SU(N_c') \times SU(\widetilde{N}_c'')$ magnetic theory}

Let us take 
the mass deformation by moving the $(N_c'-N_c)$ D4-branes suspending
between
the $NS5_L'$-brane and the NS5-brane to $+v$ direction
in the electric theory  and then
describe the Seiberg dual for the third gauge group.

$\bullet$ Mass deformation by $G$ and $\widetilde{G}$

Let us consider other magnetic theory for the same undeformed electric
theory(characterized by Figure 4A which is the same as Figure 2A)
given in the subsection 2.1 by considering the 
different deformation in electric theory(characterized by Figure 4B).
Displacing the two NS5'-branes relative each other in the $+v$
direction, as before, corresponds to turning on a quadratic
mass-deformed electric superpotential
for the field $G$ with different contractions as follows:
\bea
W_{def} = m G \widetilde{G} (\equiv m \tr \Phi'), \qquad m = 
\frac{\Delta x}{\ell_s^2}
\label{Mass1}
\eea
where 
the third gauge group indices in $G$ and $\widetilde{G}$ 
are contracted and the mass $m$ is related to the relative 
displacement $\Delta x$ between two NS5'-branes.
The gauge-singlet $\Phi'$, which has two free indices on the gauge
group
$SU(N_c')$, 
is in the 
adjoint representation for the second  gauge group $SU(N_c')$, 
i.e., ${(\bf 1, (N_c'-N_c)^2-1, 1)  \oplus (1,1,1)}$ 
under the gauge group where the gauge group $SU(N_c')$ is broken to $SU(N_c'-N_c)$. 
That is, the gauge singlet $\Phi'$ is a $(N_c'-N_c) \times (N_c'-N_c)$ matrix.
The $NS5_L'$-brane together with $(N_c'-N_c)$-color D4-branes 
is moving to the $+v$ direction  for
fixed other branes during this particular 
mass deformation(and their mirrors to
$-v$ direction). 
Then the $x^5$ coordinate 
of $NS5_R'$-brane is equal to
zero
and the $x^5$ coordinate of $NS5_L'$-branes is given by 
$ \Delta x$.
Giving an expectation value to the meson field $\Phi'$
corresponds to recombination of $N_c'$- and $N_c''$- color 
D4-branes, which will become $N_c''$-color D4-branes
in Figure 4A such that they are suspending between 
the $NS5_L'$-brane and the $NS5_R'$-brane 
and moving them into the $w$
direction. We assume, as before, that the number of colors satisfies
\bea
N_c' \geq N_c'' \geq N_c.
\nonu
\eea

Now 
we draw this ${\cal N}=1$ supersymmetric electric 
brane configuration in Figure 4B for nonvanishing mass
for the fields $G$ and $\widetilde{G}$. 
The geometric configuration for three NS-branes in Figure 4B 
is exactly the same as the first
three NS-branes in Figure 3B of \cite{Ahn07-6}.

\begin{figure}[ht]
   \epsfxsize=5.0in 
\centerline{\epsffile{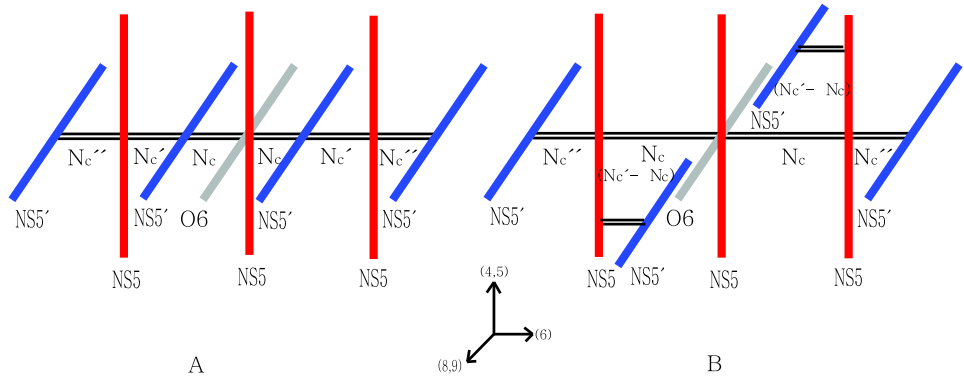}}
   \caption[FIG. \arabic{figure}.]{ 
The  ${\cal N}=1$ supersymmetric 
electric brane configuration for the gauge group $SU(N_c) \times
SU(N_c') \times SU(N_c'')$ 
and  a symmetric tensor $S$, a conjugate symmetric tensor $\widetilde{S}$,
bifundamentals $F, \widetilde{F}, G$ and $\widetilde{G}$  
with vanishing(4A) which is identical to Figure 1A and 
nonvanishing(4B) mass
for the bifundamentals $G$ and $\widetilde{G}$. 
In 4B, the $NS5_L'$-brane together with $(N_c'-N_c)$ D4-branes 
is moving to $+v$ direction(and their mirrors to $-v$ direction). 
The remaining $N_c$ D4-branes are
reconnecting with those D4-branes in 4A and finally combined
those D4-branes are suspending between NS5-brane and its mirror 
in 4B.  }
\end{figure}

$\bullet$ Seiberg dual

By applying the Seiberg dual to the third gauge group $SU(N_c'')$ factor in 
(\ref{gag}), the $NS5_{L,R}'$-branes can be located at the
inside of the two outer NS5-branes, as in Figure 5.
Starting from Figure 4B and 
interchanging the NS5-brane and the $NS5_R'$-brane(and their mirrors),
one obtains the Figure 5A with D4- and $\overline{D4}$-branes.
Before arriving at the Figure 5A, there exists an intermediate 
step where 
the $(N_c'-N_c)$ D4-branes are connecting between the 
$NS5_L'$-brane and the  $NS5_R'$-brane,  
$(N_c'-N_c'')$ D4-branes are connecting between the  $NS5_R'$-brane and   
NS5-brane(and their mirrors) as well as $N_c$ D4-branes between
$NS5_R'$-brane and its mirror.
By reconnecting the $(N_c'-N_c)$ D4-branes 
connecting between the 
$NS5_L'$-brane and the  $NS5_R'$-brane
with  
the $(N_c'-N_c)$ D4-branes connecting between  
$NS5_R'$-brane 
and the NS5-brane where we introduce $-N_c$ D4-branes and
$-N_c$ anti D4-branes and moving those combined 
D4-branes
to $+v$-direction(and their mirrors to $-v$ direction), 
one gets the final Figure 5A where we are left with 
$(N_c''-N_c)$ 
anti-D4-branes between the $NS5_R'$-brane and   
the NS5-brane.

\begin{figure}[ht]
   \epsfxsize=5.0in 
\centerline{\epsffile{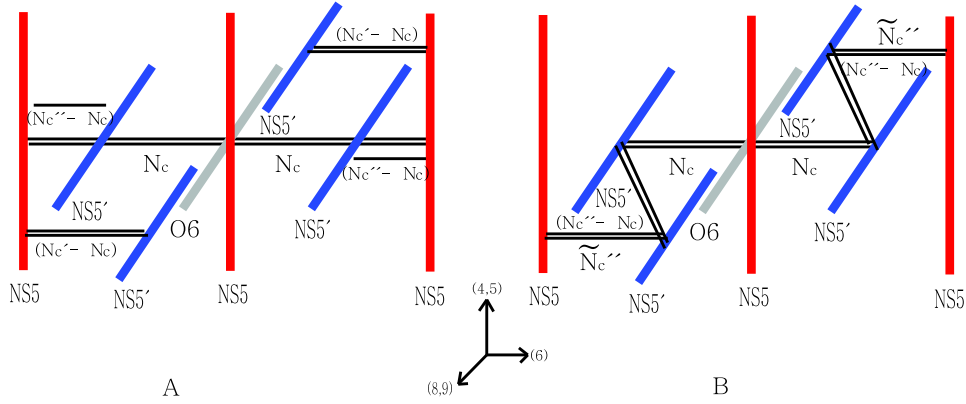}}
   \caption[FIG. \arabic{figure}.]{ 
The  ${\cal N}=1$ 
magnetic brane configuration for the gauge group $SU(N_c) \times 
SU(N_c')
\times SU(\widetilde{N}_c''=N_c'-N_c'')$ 
with D4-
and $\overline{D4}$-branes(5A) and with 
a misalignment between D4-branes(5B) when the NS5'-branes of 5A are close to
each other. The number of tilted D4-branes can be written as $N_c''-N_c=
(N_c'-N_c)-\widetilde{N}_c''$ in 5B.  }
\end{figure}

When two NS5'-branes in Figure 5A are close to each other, then 
it leads to Figure 5B
by realizing that the number of $(N_c'-N_c)$
D4-branes connecting between $NS5_L'$-brane and NS5-brane in Figure
5A can
be rewritten as $(N_c''-N_c)$ plus $\widetilde{N}_c''$.
The brane configuration consisting of NS5-brane and two NS5'-branes in
Figure 5B is
exactly the same as those in Figure 4B of \cite{Ahn07-6}. 
Moreover if we change the $O6^{+}$-plane into the $O6^{-}$-plane,
replace the number of color $N_c$ by $2N_c$, and  remove the middle
NS5-brane 
in Figure 5, 
then this is exactly the same as the Figure 14 of \cite{Ahn07-6}.

The dual gauge group can be read off and is given by
\bea
SU(N_c) \times SU(N_c') \times SU(\widetilde{N}_c''=N_c'-N_c''). 
\label{dualneww}
\eea
The matter contents are 
 a symmetric tensor field $S$ 
charged under $({\bf \frac{1}{2}
N_c(N_c+1), 1, 1})$,
the field $F$ 
 charged under
$({\bf N_c, \overline{N_c'}, 1})$, a field $g$ charged under
$({\bf 1, N_c', \overline{\widetilde{N}_c''}})$ 
 and their conjugates 
$\widetilde{S}, \widetilde{F}$ and $\widetilde{g}$
under the dual gauge group
(\ref{dualneww})
and  
the gauge-singlet $\Phi'$ which is in the 
adjoint representation for the second dual gauge group, 
in other words,   
$ ({ \bf   1,  (N_c'-N_c)^2-1,1})  \oplus  ({\bf 1,1,1})$ under the 
dual gauge group (\ref{dualneww})  where the gauge group $SU(N_c')$ is 
broken to $SU(N_c'-N_c)$.
Then the $\Phi'$ is a $(N_c'-N_c) \times (N_c'-N_c)$ matrix.
Only $(N_c'-N_c)$ D4-branes among $N_c'$ D4-branes 
can participate in the mass deformation
we are considering.

$\bullet$ Meta-stable brane configuration

The cubic superpotential plus the mass term
in the magnetic theory is given by
\bea
W_{dual} = \Phi' g \widetilde{g} + m \tr \Phi'
\label{superpo11}
\eea
where $\Phi'$ was defined as $\Phi' \equiv G \widetilde{G}$ from (\ref{Mass1}) and 
the third gauge group indices in $G$ and $\widetilde{G}$ 
are contracted and each second gauge group index in them is encoded in 
$\Phi'$. Although the $\Phi'$ that has second gauge group indices 
looks similar to the previous 
$\Phi''$ that has third gauge group indices, 
their group indices are different from each other. 
Here the magnetic fields $g$ and $\widetilde{g}$  
correspond to 4-4 strings connecting 
the $\widetilde{N}_c''$-color D4-branes(that are 
connecting between the $NS5_L'$-brane
and the NS5-brane in Figure 5B) with $N_c'$-flavor 
D4-branes.
Among these $N_c'$-flavor D4-branes, only the strings ending on
the upper $(N_c'-N_c'')$ D4-branes and 
on the tilted  $(N_c''-N_c)$ 
D4-branes in Figure 5B enter the cubic superpotential term. 
Note that the summation of these D4-branes is equal to $(N_c'-N_c)$.
Although the $(N_c'-N_c)$ D4-branes in Figure 5A cannot move any
directions,
the tilted $(N_c''-N_c)$-flavor D4-branes can move $w$ direction.
The remaining upper $\widetilde{N}_c''$ D4-branes are fixed also and cannot 
move any direction. 
As before, it is useful to understand the transition from Figure 5A to
Figure 5B to note that 
there is a decomposition 
\bea
(N_c'-N_c)=(N_c''-N_c)+\widetilde{N}_c''.
\nonu
\eea 

The brane configuration for zero mass for the bifundamental,
which has only a cubic superpotential (\ref{superpo11}),
can be obtained from Figure 5A by moving
the upper  NS5'-brane together with $(N_c'-N_c)$ color D4-branes 
into the origin $v=0$(and their mirrors).
Then the number of dual colors for D4-branes 
becomes $N_c$ between the $NS5_L'$-brane and its mirror, 
$N_c'$ between the $NS5_L'$-brane and the  $NS5_R'$-brane
and $\widetilde{N}_c''$ 
between $NS5_R'$-brane and NS5-brane.
Or starting from Figure 4A and moving the 
NS5-brane to the right all the
way past the $NS5_R'$-brane(and their mirrors),
one also obtains the corresponding magnetic brane configuration
for massless case.

The brane configuration in Figure 5A is stable as long as the
distance $\Delta x$ between the upper NS5'-brane and 
the lower NS5'-brane is large. 
If they are close to each other, then this brane
configuration is unstable to decay  to 
the meta-stable brane configuration in Figure
5B.
One can regard these brane configurations as particular states in the
magnetic gauge theory with the gauge group (\ref{dualneww}) and
superpotential (\ref{superpo11}).
Then the  upper tilted $(N_c'-N_c-\widetilde{N}_c'')$ flavor D4-branes of 
straight brane configuration
of
Figure 5B can bend due to the fact that there exists an attractive
gravitational interaction
between those flavor D4-branes and NS5-brane, 
as long as $y_1$ is very large.
Of course, their mirrors, the lower tilted 
$(N_c'-N_c-\widetilde{N}_c'')$ flavor D4-branes of 
straight brane configuration
of
Figure 5B can bend and their trajectories connecting 
two NS5'-branes should be preserved under the $O6^{+}$-plane action.
The $N_c$ D4-branes between the $NS5_R'$-brane and its mirror can
slide along the $w$ direction.

When the upper NS5'-brane(or $NS5_L'$-brane) 
is replaced by coincident $(N_c'-N_c)$ 
D6-branes in Figure 5B, this brane configuration looks similar to the
Figure 7 
found in \cite{Ahn07-4}. 
The only difference between these two brane configurations 
appears the brane contents at the origin $x^6=0$.
The gauge group in \cite{Ahn07-4} was   
given by 
$SU(n_c) \times SU(n_f'+n_c-n_c')$ 
with $n_f'$ 
fundamentals, bifundamentals, an antisymmetric flavor, a
conjugate symmetric flavor, and various gauge singlets where there are NS5'-brane,
$O6^{\pm}$-planes,
and eight semi infinite D6-branes at $x^6=0$. 
Then the our $N_c$ corresponds to the $n_c$,
the number $(N_c'-N_c)$ corresponds to $n_f'$,
and 
moreover $N_c''$ corresponds to the $n_c'$ of \cite{Ahn07-4}. 

$\bullet$ Gauge theory at small $\Delta x$ 

The low energy dynamics of the magnetic brane configuration 
can be described by the ${\cal N}=1$ supersymmetric gauge theory
with gauge group (\ref{dualneww})
and the gauge couplings for the three gauge group factors are
given by similarly.
The dual gauge theory has  a meson  $\Phi'$  and 
bifundamentals $f, \widetilde{f}, g$ and 
$\widetilde{g}$ 
 as well as 
a symmetric tensor $S$ and its conjugate $\widetilde{S}$ 
under the dual gauge
group (\ref{dualneww}) and the superpotential 
corresponding to the Figures 5A and 5B is given by 
\bea
W_{dual} = h \Phi' g \widetilde{g} - h \mu^2 \tr \Phi', \qquad h^2 = g_{2,
  mag}^2.
\label{dualw}
\eea
Then $ g \widetilde{g}$ is a $\widetilde{N}_c'' \times \widetilde{N}_c''$ 
matrix where the second gauge group indices for $g$ and $\widetilde{g}$ 
are contracted with those
of $\Phi'$ while the  $\Phi'$ is a 
$(N_c'-N_c) \times (N_c'-N_c)$ matrix and the $\mu^2$ is the same as before.
The product $g \widetilde{g}$ has the same representation for the 
product of quarks
and 
the second gauge group indices for the field $\Phi'$ play the
role of the flavor indices, as before.

Therefore, the F-term equation, the derivative of $W_{dual}$
(\ref{dualw}) 
with respect to the
meson field $\Phi'$ cannot be satisfied if the $(N_c'-N_c)$ exceeds
$\widetilde{N}_c''$.
So the supersymmetry is broken.   
That is, 
there exist three equations from F-term conditions:
$
g^a \widetilde{g}_b -\mu^2 \delta^a_b =0$ and $ \Phi' g =0=\widetilde{g} \Phi'$.
Then the solutions for these
are given by 
\bea
<g>   = 
\left(
\begin{array}{c}
\mu  {\bf 1}_{\widetilde{N}_c''}  \\
0
\end{array}
\right), 
\qquad
<\widetilde{g}>   = 
\left(
\begin{array}{cc}
\mu  {\bf 1}_{\widetilde{N}_c''} & 0  \\
\end{array}
\right), 
\qquad
<\Phi'> =
 \left(
\begin{array}{cc}
0  & 0  \\
0 & \Phi_0'  {\bf 1}_{(N_c'-N_c)-\widetilde{N}_c''} 
\end{array}
\right) 
\label{point12-1}
\eea
where the zero of $<g>$ is a $
(N_c'-N_c-\widetilde{N}_c'') \times \widetilde{N}_c''$ 
matrix, the zero of $<\widetilde{g}>$ is a
$\widetilde{N}_c'' \times (N_c'-N_c-\widetilde{N}_c'') $ matrix and 
the zeros of $<\Phi'>$ are $\widetilde{N}_c'' \times \widetilde{N}_c''$,
$\widetilde{N}_c'' \times 
(N_c'-N_c-\widetilde{N}_c'')$ and $(N_c'-N_c-\widetilde{N}_c'') \times
\widetilde{N}_c''$ 
matrices.
Then one can expand these fields around on a point (\ref{point12-1}), as
in \cite{ISS,Shih} and one arrives at the relevant superpotential
up to quadratic order in the fluctuation. 
At one loop, the effective potential $V_{eff}^{(1)}$ for $\Phi'_0$
leads to the positive value for $m_{\Phi'_0}^2$ implying that these
vacua are stable.
The gauge theory analysis 
is only valid in the regime where 
$\Delta x$ is smaller than $\exp(-\frac{C}{g_s})$ with some positive
constant $C$. 

\subsection{
${\cal N}=1$ 
$SU(\widetilde{N}_c) \times SU(N_c') \times SU(N_c'')$ magnetic theory}

Let us consider the Seiberg dual for the first gauge group and take 
the mass deformation by moving the $(N_c'-N_c'')$ D4-branes between 
the middle NS5-brane and the $NS5_M'$-brane to $+v$ direction.

$\bullet$ Seiberg dual and mass deformation

Starting from the Figure 4A, we apply the Seiberg dual to the first
gauge group 
$SU(N_c)$ factor and the $NS5_L'$-brane and its mirror are 
interchanged each other. 
Then the brane configuration 
\footnote{This duality symmetry in field theory side was tested in 
\cite{ILS} and the brane configuration in type IIA string theory 
for the supersymmetric magnetic
theory was studied in \cite{LLL}. Moreover, the nonsupersymmetric
meta-stable brane configuration was found in \cite{Ahn07} where the
correct movement of the branes were crucial.  }
is the same as Figure 4A except that 
the number of color $\widetilde{N}_c$
is given by $\widetilde{N}_c=2N_c'-N_c$ from \cite{Ahn07-4,Ahn07}.
By rotating the NS5-brane by $\frac{\pi}{2}$(leading to
$NS5_M'$-brane) and moving
it  together with $(N_c'-N_c'')$ D4-branes
to $+ v$ direction, 
then the $(N_c'-N_c'')$ 
D4-branes are connecting between the 
$NS5_L'$-brane and the $NS5_M'$-brane and 
$\widetilde{N}_c$ D4-branes connecting between the middle NS5-brane and   
$NS5_L'$-brane as well as $N_c''$ D4-branes between the $NS5_L'$-brane
and the $NS5_R'$-brane(and their mirrors). 

By introducing $(N_c'-N_c'')$ D4-branes and $(N_c'-N_c'')$ 
anti-D4-branes  between the middle NS5-brane and   
$NS5_L'$-brane, reconnecting the former with  
the $(N_c'-N_c'')$ D4-branes connecting between the $NS5_L'$-brane and
the $NS5_M'$-brane 
and moving those combined D4-branes
to $+v$-direction(and their mirrors to $-v$ direction), 
one gets the final Figure 6A where we are left with 
$(N_c'-N_c''-\widetilde{N}_c)$ anti-D4-branes between the middle NS5-brane and   
$NS5_L'$-brane.

Now 
we draw this ${\cal N}=1$ supersymmetric magnetic
brane configuration in Figure 6.
We assume, as before, that the number of colors satisfies
\bea
2N_c' \geq N_c \geq N_c' +N_c''.
\nonu
\eea

\begin{figure}[ht]
   \epsfxsize=5.0in 
\centerline{\epsffile{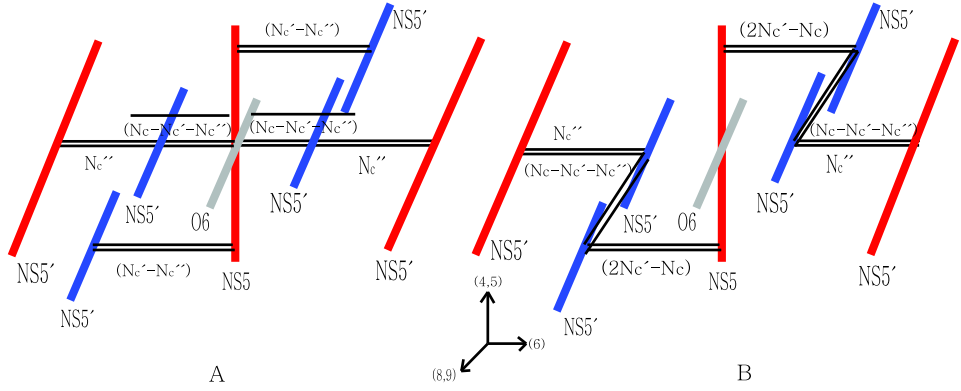}}
   \caption[FIG. \arabic{figure}.]{ 
The  ${\cal N}=1$ magnetic brane configuration for the gauge group 
$SU(\widetilde{N}_c=2N_c'-N_c) 
\times SU(N_c') \times SU(N_c'')$ 
with D4-
and $\overline{D4}$-branes(6A) and with 
a misalignment between D4-branes(6B) when the NS5'-branes are close to
each other. The number of tilted D4-branes in 6B can be written as
$N_c-N_c'-N_c''=(N_c'-N_c'')-\widetilde{N}_c$.   }
\end{figure}

When two NS5'-branes in Figure 6A are close to each other, then 
it leads to Figure 6B
by realizing that the number of $(N_c'-N_c'')$
D4-branes connecting between the NS5-brane and the $NS5_M'$-brane in Figure
6A can
be rewritten as $(N_c-N_c'-N_c'')$ plus $\widetilde{N}_c$.
The brane configuration consisting of the middle NS5-brane and first 
two NS5'-branes in
Figure 6B is
exactly the same as the first three NS-branes in 
Figure 2B of \cite{Ahn07-6}. 

The dual gauge group is given by
\bea
SU(\widetilde{N}_c=2N_c'-N_c) \times SU(N_c') \times SU(N_c'')
\label{dual2}
\eea
where the number of dual color can be obtained from the linking number 
counting, as done in \cite{Ahn07-4,Ahn07}. Note that the number of
flavors of D6-branes in \cite{Ahn07-4} is zero because in the present
case, there are no D6-branes.
The matter contents are the flavor singlet $f$ in the bifundamental 
representation $({\bf \widetilde{N}_c, \overline{N_c'}, 1})$
and its complex conjugate field $\widetilde{f}$ in the bifundamental 
representation  $({\bf \overline{\widetilde{N}_c}, N_c', 1})$,
under the dual gauge
group (\ref{dual2}) 
and  the gauge singlet $\Phi'$ in the representation for 
$({\bf 1,(N_c'-N_c'')^2-1, 1}) \oplus ({\bf 1, 1, 1})$ under 
the dual gauge group  where the gauge group $SU(N_c')$ is 
broken to $SU(N_c'-N_c'')$.
There are also
the symmetric flavor $s$ for $SU(\widetilde{N}_c)$ and the conjugate 
symmetric flavor $\widetilde{s}$ for $SU(\widetilde{N}_c)$ as well as
$G$ and $\widetilde{G}$.
Then the $\Phi'$ is a $(N_c'-N_c'') \times (N_c'-N_c'')$ matrix.
Only $(N_c'-N_c'')$ D4-branes among $N_c'$ D4-branes 
can participate in the mass deformation.
A cubic superpotential is an interaction between dual ``quarks''
and a meson. 

$\bullet$ Meta-stable brane configuration

Then the dual magnetic superpotential, by adding the mass term
for the bifundamental $F$ which can be 
interpreted as a linear term in the meson $\Phi'$ to this cubic
superpotential, is given by
\bea
W_{dual} =  \Phi' f \widetilde{f} + m \tr \Phi', \qquad \Phi' \equiv F
\widetilde{F},
\qquad m = \frac{\Delta x}{\ell_s^2}
\label{superpo1}
\eea
where $\Phi'$ was defined as $\Phi' \equiv F \widetilde{F}$  and 
the first gauge group indices in $F$ and $\widetilde{F}$ 
are contracted and each second gauge group index in them is encoded in 
$\Phi'$.

Here the magnetic fields $f$ and $\widetilde{f}$  
correspond to 4-4 strings connecting 
the $\widetilde{N}_c$-color D4-branes(that are 
connecting between the NS5-brane
and the $NS5_M'$-brane in Figure 6B) with $N_c'$-flavor 
D4-branes.
Among these $N_c'$-flavor D4-branes, only the strings ending on
the upper $(2N_c'-N_c)$ D4-branes and 
on the tilted  $(N_c-N_c'-N_c'')$ 
D4-branes in Figure 6B enter the cubic superpotential term. 
Note that the summation of these D4-branes is equal to $(N_c'-N_c'')$.
Although the $(N_c'-N_c'')$ D4-branes in Figure 6A cannot move any
directions,
the tilted $(N_c-N_c'-N_c'')$-flavor D4-branes can move $w$ direction.
The remaining upper $\widetilde{N}_c$ D4-branes are fixed also and cannot 
move any direction. 
It is useful to understand the transition from Figure 6A to
Figure 6B to note that 
there is a decomposition 
\bea
(N_c'-N_c'')=(N_c-N_c'-N_c'')+\widetilde{N}_c.
\nonu
\eea 

The brane configuration for zero mass for the bifundamentals
can be obtained from Figure 6A by  pushing the $NS5_M'$-brane  
with $(N_c'-N_c'')$ D4-branes into the origin $v=0$.
Then the number of dual colors for D4-branes 
becomes 
$\widetilde{N}_c$ between the NS5-brane and the $NS5_L'$-brane,
$N_c'$ between the $NS5_L'$-branes and the $NS5_M'$-brane, and 
$N_c''$ between the $NS5_M'$-brane and the $NS5_R'$-brane. 

The brane configuration in Figure 6A is stable as long as the
distance $\Delta x$ between the upper $NS5_M'$-brane and 
the middle $NS5_L'$-brane is large. 
If they are close to each other, then this brane
configuration is unstable to decay and leads to 
the meta-stable brane configuration in Figure
6B.
One can regard these brane configurations as particular states in the
magnetic gauge theory with the gauge group (\ref{dual2}) and
superpotential (\ref{superpo1}).

When the $NS5_M'$-brane which is connected by $\widetilde{N}_c$
D4-branes 
is replaced by $(N_c'-N_c'')$
D6-branes,
the brane configuration of Figure 6B together with the rotation of 
$NS5_R'$-brane by $\frac{\pi}{2}$ is the same as the Figure 4(with the
movement of $n_f$ D6-branes to the opposite $v$ direction) studied in 
\cite{Ahn07-4}
 where the gauge group is   
given by 
$SU(2n_f+2n_c'-n_c) \times SU(n_c')$ 
with $n_f$ fundamentals, bifundamentals, a symmetric flavor, a
conjugate symmetric flavor, and various gauge singlets. 
Then the our $N_c$ corresponds to the $n_c$,
the number $(N_c'-N_c'')$ corresponds to $n_f$,
and 
our $N_c''$ corresponds to the $n_c'$ of \cite{Ahn07-4}. 

One can perform similar analysis in our brane configuration 
since one can take into account the behavior of
parameters geometrically in the presence of $O6^{+}$-plane.
Then the upper tilted $(N_c'-N_c''-\widetilde{N}_c)$ flavor D4-branes of 
straight brane configuration
of
Figure 6B bend due to the fact that there exists an attractive
gravitational interaction
between those flavor D4-branes and NS5-brane from the DBI action. 
Of course, their mirrors, the lower 
$(N_c'-N_c''-\widetilde{N}_c)$ flavor D4-branes of 
straight brane configuration
of
Figure 6B bend and their trajectory connecting 
two NS5'-branes should be preserved under the $O6^{+}$-plane action.
The correct choice for the ground state of the system 
depends on the parameters $y_i$ and $\Delta x$. 

The $N_c''$ D4-branes between the $NS5_L'$-brane and the $NS5_R'$-brane can
slide along the $w$ direction.

$\bullet$ Gauge theory at small $\Delta x$ 

The low energy dynamics of the magnetic brane configuration 
can be described by the ${\cal N}=1$ supersymmetric gauge theory
with gauge group (\ref{dual2})
and the gauge couplings for the  gauge group factors are
given by similarly.
The dual gauge theory has  a meson field  $\Phi'$  and 
bifundamental $f$ in the representation 
 $({\bf \widetilde{N}_c, \overline{N_c'}, 1})$ under the dual gauge
group (\ref{dual2}) and the superpotential 
corresponding to the Figures 6A and 6B is given by
\bea
W_{dual} = h \Phi' f \widetilde{f} - h \mu^2 \tr \Phi', \qquad
h^2=  g_{2,mag}^2 
\nonu
\eea
and the  mass parameter $\mu^2$ is the same as before.
See also the relevant papers \cite{Ahn07,Ahn07-5}.
Then $ f \widetilde{f}$ is 
a $\widetilde{N}_c \times \widetilde{N}_c$ 
matrix where the second gauge group indices for $f$ and $\widetilde{f}$ 
are contracted with those
of $\Phi'$ while $\Phi'$ is a 
$(N_c'-N_c'') \times (N_c'-N_c'')$ matrix.
Although the field $f$ itself is an antifundamental in the second gauge
group
which is a different feature, compared with the singlet 
representation for the usual quark
coming from D6-branes \cite{Ahn07-4},
the product $f \widetilde{f}$ has the same representation with the 
product of quarks.
Moreover, the second gauge group indices for the field $\Phi'$ play the
role of the flavor indices.

Therefore, the F-term equation, the derivative of $W_{dual}$ with respect to the
meson field $\Phi'$ cannot be satisfied if the $(N_c'-N_c'')$ exceeds
$\widetilde{N}_c$.
So the supersymmetry is broken.   
That is, 
there are three equations from F-term conditions:
$
f^a \widetilde{f}_b -\mu^2 \delta^a_b =0, \Phi' f =0$, and $\widetilde{f} \Phi'=0$.
Then the solutions for these
are given by 
\bea
<f>   = 
\left(
\begin{array}{c}
\mu   {\bf 1}_{\widetilde{N}_c}  \nonu \\
0
\end{array}
\right), \quad
<\widetilde{f}>   = 
\left(
\begin{array}{cc}
\mu   {\bf 1}_{\widetilde{N}_c} & 0 \nonu \\
\end{array}
\right), 
\quad
<\Phi'> =
 \left(
\begin{array}{cc}
0  & 0
 \\
0 & \Phi_0'  {\bf 1}_{(N_c'-N_c''-\widetilde{N}_c)} 
\end{array}
\right).
\nonu
\eea
At one loop, the effective potential $V_{eff}^{(1)}$ for $\Phi_0'$
leads to the positive value for $m_{\Phi_0'}^2$ implying that these
vacua are stable.
The gauge theory analysis where the theory will be strongly coupled in
the IR region $N_c' -N_c''> 2\widetilde{N}_c-2$ is only valid in the regime where 
$\Delta x$ is smaller than $\exp(-\frac{C}{g_s})$ with some positive
constant $C$, as done in \cite{Ahn07-5}. 

\section{Meta-stable brane configurations-II}

In this section, we continue to 
construct the meta-stable brane configurations 
for the same gauge theory we have considered in previous section but 
different matter contents.
In the brane configuration context, the roles of NS5-brane and
NS5'-brane are interchanged in an electric brane configuration  
and the brane contents at the origin $x^6=0$ are different from the
previous case as we will see later.

The type IIA brane configuration  corresponding to 
${\cal N}=1$ supersymmetric gauge theory with
gauge group
\bea
SU(N_c) \times SU(N_c') \times SU(N_c'') 
\label{gag1}
\eea
and 
with an antisymmetric tensor field $A$ charged under $({\bf \frac{1}{2}
N_c(N_c-1), 1, 1})$, 
 a conjugate symmetric tensor field $\widetilde{S}$ 
charged under $({\bf \overline{\frac{1}{2}
N_c(N_c+1)}, 1, 1})$,  an eight fundamentals $\hat{Q}$ charged under
 $({\bf N_c, 1, 1})$,
a field $F$ charged under
$({\bf N_c, \overline{N_c'}, 1})$, a field $G$ charged under
$({\bf 1, N_c', \overline{N_c''}})$, and their conjugates 
$\widetilde{F}$ and $\widetilde{G}$ 
can be described by 
the left $NS5_L$-brane, 
the  
NS5'-brane, the right $NS5_R$-brane(and their mirrors),
 $N_c$-, $N_c'$-  and $N_c''$-color D4-branes(and their mirrors) as well as
$O6^{+}$-plane(0123789), $O6^{-}$-plane(0123789), eight 
half D6-branes(0123789) and the middle NS5'-brane
\footnote{When we say about NS-branes(NS5-brane or
  NS5'-brane) in this section, 
those NS-branes are located at $x^6 >0$. 
We assume that their mirrors behave according to ${\bf Z}_2$ symmetry
of $O6^{+}$-plane. 
That is, there exist three NS-branes:
$NS5_L$-brane,
NS5'-brane and $NS5_R$-brane from Figure 6A.}.
See also the relevant works \cite{LLL1,BHKL,EGKT} which have dealt with the
gauge theory realized by the first factor in (\ref{gag1}) with an
antisymmetric tensor, a conjugate symmetric tensor, fundamentals and 
antifundamentals. 

Let us place an $O6^{+}$-plane at the origin $x^6=0$
and let us denote the $x^6$ 
coordinates for the $NS5_L$-brane, the NS5'-brane and the $NS5_R$-brane 
by $x^6=y_1, y_1+y_2, y_1+y_2+y_3$
respectively. Their mirrors can be understood similarly.
The $N_c$ D4-branes 
are suspended between the 
$NS5_L$-brane and its mirror, 
the $N_c'$ D4-branes 
are suspended between the 
$NS5_L$-brane and the NS5'-brane(and their mirrors), and 
the $N_c''$ D4-branes  
are suspended between the NS5'-brane and the $NS5_R$-brane(and their mirrors).
The fields $A$ and $\widetilde{S}$  correspond to 4-4 strings connecting 
the $N_c$-color D4-branes with $x^6 < 0$ with $N_c$-color D4-branes
with $x^6 > 0$.
We draw this brane configuration in Figure 7A \footnote{According to
the observation of \cite{LLL1,BHKL,EGKT}, this ``fork'' 
brane configuration
contains the NS5'-brane embedded in an O6-plane at $x^7=0$. This 
NS5'-brane divides the O6-plane into two separated regions corresponding
to positive $x^7$ and negative $x^7$. Then RR charge of the O6-plane
jumps from $-4$ to $+4$. Furthermore, eight semi-infinite D6-branes
are present in the positive $x^7$ region. This is necessary for the
vanishing of the six dimensional anomaly.  Further discussions on the
gauge symmetry or flavor symmetry of this brane configuration 
can be found in \cite{GK98} or
\cite{Ahn07-1}.
 } for the vanishing mass
for the bifundamentals.

Let us discuss three different magnetic gauge theories by taking each
mass deformation.

\subsection{${\cal N}=1$ 
$SU(\widetilde{N}_c) \times SU(N_c') \times SU(N_c'')$ magnetic theory}

Let us take 
the mass deformation by moving the $(N_c'-N_c'')$ D4-branes between 
the $NS5_L$-brane and the NS5'-brane to $+v$ direction
in the electric theory and  consider the Seiberg dual for the first gauge group. 

$\bullet$ Mass deformation by $F$ and $\widetilde{F}$
 
There is no electric
superpotential corresponding to the Figure 7A. 
It is known in \cite{Ahn07-4} that the classical superpotential 
without $NS5_R$-brane(and its mirror) for the particular orientation
for NS-branes in Figure 7A is vanishing.
Now we are adding the $NS5_R$-brane(and its mirrors) 
into the brane configuration of 
\cite{Ahn07-4} together with $N_c''$ D4-branes.
Since the angle between the NS5'-brane and the extra $NS5_R$-brane
is $\frac{\pi}{2}$ in Figure 7A, the mass for the extra adjoint field $\mu''$
corresponding to the
gauge group $SU(N_c'')$
becomes large and after integrating out this adjoint field, then
the extra classical
superpotential term which has a factor $\frac{1}{\mu''}$ will vanish.  

\begin{figure}[ht]
   \epsfxsize=5.0in 
\centerline{\epsffile{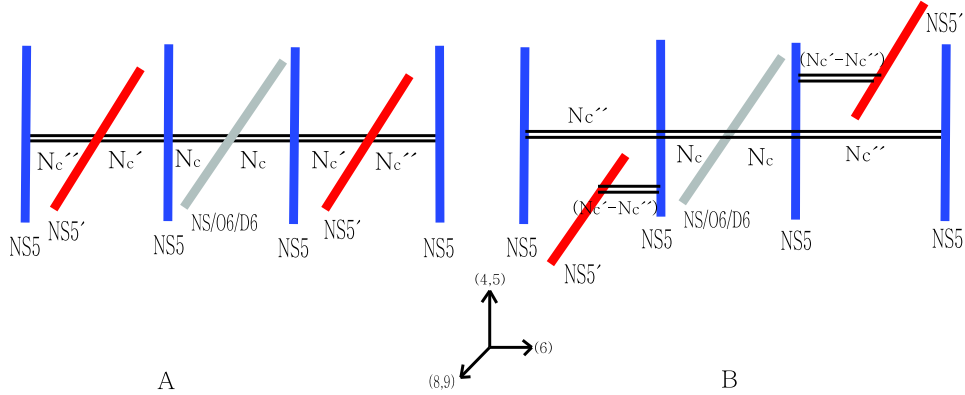}}
   \caption[FIG. \arabic{figure}.]{ 
The  
  ${\cal N}=1$ supersymmetric 
electric brane configuration for the gauge group $SU(N_c) \times
SU(N_c') \times SU(N_c'')$ 
and an antisymmetric tensor $A$, a conjugate symmetric tensor $\widetilde{S}$,
bifundamentals $F, \widetilde{F}, G, \widetilde{G}$ and fundamentals
  $\hat{Q}$  
with vanishing(7A) and 
nonvanishing(7B) mass
for the bifundamentals $F$ and $\widetilde{F}$. 
The NS5'-brane with $(N_c'-N_c'')$ D4-branes is moving to $+v$  direction 
in 7B(and their mirrors to $-v$ direction) 
and the remaining $N_c''$ D4-branes are recombined with those D4-branes
connecting between NS5'-brane and $NS5_R$-brane in 7B.
The notation for
NS/O6/D6 refers to a combination of a middle
NS5'-brane, $O6^{+}$-plane with $x^7 \leq 0$, 
$O6^{-}$-plane with $x^7 \geq 0$ and eight half D6-branes with $x^7
  \geq 0$.  }
\end{figure}

Now let us deform 
this theory. 
Displacing the two NS5'-branes relative each other in the $+v$ 
direction corresponds to turning on a quadratic
superpotential
for the bifundamentals $F$ and $\widetilde{F}$
where the $\Phi'$
is a meson field
\bea
W_{def} = m F \widetilde{F} (\equiv m \tr \Phi'), \qquad m = 
\frac{\Delta x}{\ell_s^2}
\nonu
\eea
where 
the first gauge group indices in $F$ and $\widetilde{F}$ 
are contracted and the mass $m$ is related to the relative 
displacement $\Delta x$ between the middle NS5'-brane and the NS5'-brane.
The gauge-singlet $\Phi'$, which has two free indices on the gauge
group
$SU(N_c')$, for the  gauge group 
is in the 
adjoint representation for the second  gauge group $SU(N_c')$, 
i.e., ${(\bf 1, (N_c'-N_c'')^2-1, 1)  \oplus (1,1,1)}$ 
under the gauge group  where the gauge group $SU(N_c')$ is broken to 
$SU(N_c'-N_c'')$. 
That is, the gauge singlet $\Phi'$ is a $(N_c'-N_c'') \times
(N_c'-N_c'')$ matrix implying that only $(N_c'-N_c'')$ D4-branes
participating in the mass deformation.
The NS5'-brane together with $(N_c'-N_c'')$-color D4-branes 
is moving to the $+v$ direction  for
fixed other branes during this particular 
mass deformation(and their mirrors to
$-v$ direction). 
We assume that the number of colors satisfies
\bea
2N_c' \geq N_c-4 \geq N_c'+N_c''.
\nonu
\eea

We draw this ${\cal N}=1$ supersymmetric 
brane configuration in Figure 7B for nonvanishing mass
for the bifundamentals
by moving the NS5'-brane with 
$(N_c'-N_c'')$ color D4-branes 
to the $+v$ direction and their mirrors to $-v$
direction. 
The geometric configuration for three NS-branes in Figure 7B 
is exactly the same as the first
three NS-branes in Figure 1B of \cite{Ahn07-6}.

$\bullet$ Seiberg dual

Let us apply the Seiberg dual to the first gauge group $SU(N_c)$ factor.
Starting from Figure 7B and moving the $NS5_L$-brane to the left all the
way past the middle NS5'-brane(and the mirror of $NS5_L$-brane to the
right of the middle NS5'-brane),
one obtains the Figure 8A.
Before arriving at the Figure 8A, there exists an intermediate 
step where 
the $\widetilde{N}_c$ D4-branes are connecting between the 
$NS5_L$-brane and its mirror,  
$(N_c'-N_c'')$ D4-branes are connecting between the  $NS5_L$-brane and   
NS5'-brane(and their mirrors) as well as $N_c''$ D4-branes between
NS5'-brane and the $NS5_R$-brane.
By introducing $(N_c'-N_c'')$ D4-branes and $(N_c'-N_c'')$ 
anti-D4-branes  between $NS5_L$-brane and   
its mirror, 
one gets the final Figure 8A where we are left with 
$(N_c'-N_c''-\widetilde{N}_c)$ 
anti-D4-branes between the $NS5_L$-brane and   
its mirror.

When two NS5'-branes in Figure 8A are close to each other, it becomes 
the meta-stable brane configuration Figure 8B
 by realizing that the number of $(N_c'-N_c'')$
D4-branes connecting between $NS5_L$-brane and NS5'-brane can
be rewritten as $(N_c-N_c'-N_c''-4)$ plus $\widetilde{N}_c$. 

\begin{figure}[ht]
   \epsfxsize=5.0in 
\centerline{\epsffile{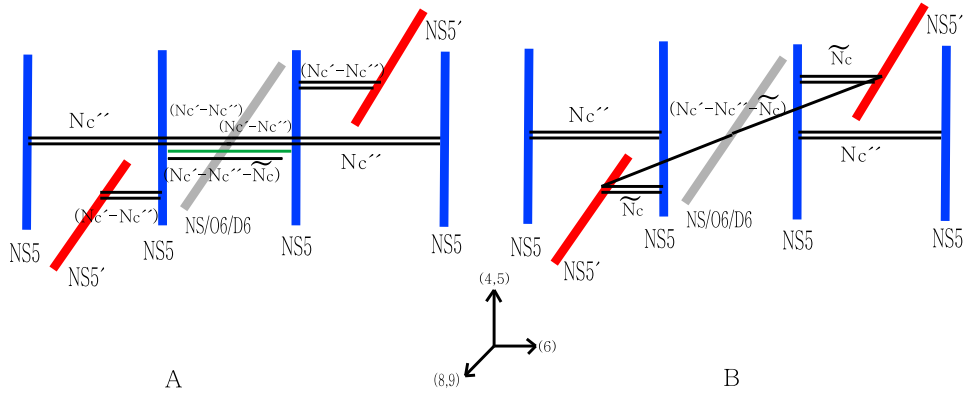}}
   \caption[FIG. \arabic{figure}.]{ 
The  ${\cal N}=1$ magnetic brane configuration for the gauge group 
$SU(\widetilde{N}_c=2N_c'-N_c+4) 
\times SU(N_c') \times SU(N_c'')$ corresponding to Figure 7B with D4-
and $\overline{D4}$-branes(8A) and with 
a misalignment between D4-branes(8B) when the NS5'-branes are close to
each other.
 The number of tilted D4-branes in 8B can be written as
$N_c-N_c'-N_c''-4=(N_c'-N_c'')-\widetilde{N}_c$.  }
\end{figure}

The gauge group is given by
\bea
SU(\widetilde{N}_c=2N_c'-N_c+4) \times SU(N_c') \times SU(N_c'')
\label{dualdual}
\eea
where the number of dual color 
\footnote{This duality symmetry in field theory side was tested in 
\cite{ILS} also and the brane configuration in type IIA string theory 
for the supersymmetric magnetic
theory was studied in \cite{LLL1,BHKL,EGKT}. Moreover, the nonsupersymmetric
meta-stable brane configuration was found in \cite{Ahn07-1} where the
correct movement of the branes were crucial.  }
can be obtained from the linking number 
counting, as done in \cite{Ahn07-4,Ahn07-1}.
Note that the number of
flavors of D6-branes in \cite{Ahn07-4} is zero because in the present
case, there are no D6-branes.
The matter contents are the flavor singlet $f$ in the bifundamental 
representation $({\bf \widetilde{N}_c, \overline{N_c'}, 1})$
and its complex conjugate field $\widetilde{f}$ in the bifundamental 
representation  $({\bf \overline{\widetilde{N}_c}, N_c', 1})$,
and  the gauge singlet $\Phi' \equiv F \widetilde{F}$ 
in the representation for 
$({\bf 1, (N_c'-N_c'')^2-1, 1}) \oplus ({\bf 1, 1, 1})$, under the
dual gauge 
group  where the gauge group $SU(N_c')$ is broken to $SU(N_c'-N_c'')$.
There are also
the antisymmetric flavor $a$, the conjugate 
symmetric flavor $\widetilde{s}$ and eight fundamentals $\hat{q}$ for 
$SU(\widetilde{N}_c)$ as well as $G$ and $\widetilde{G}$.

$\bullet$ Meta-stable brane configuration

Then the dual magnetic superpotential, by adding the mass term
 for the bifundamental, is given by
\bea
W_{dual} =  \Phi' f \widetilde{f}  + m \tr \Phi'
+ \hat{q}
 \widetilde{s} \hat{q}. 
\label{super}
\eea
Here the magnetic fields $f$ and $\widetilde{f}$  
correspond to 4-4 strings connecting 
the $\widetilde{N}_c$-color D4-branes(that are 
connecting between the $NS5_L$-brane
and the NS5'-brane in Figure 8B) with $N_c'$-flavor 
D4-branes(which  are realized 
as corresponding D4-branes in Figure 8A).
Among these $N_c'$-flavor D4-branes, only the strings ending on
the upper $\widetilde{N}_c$ D4-branes and 
on the tilted  $(N_c'-N_c''-\widetilde{N}_c)$ 
D4-branes in Figure 8B enter the above cubic superpotential term. 
Note that the summation of these D4-branes is equal to $(N_c'-N_c'')$.
Although the $(N_c'-N_c'')$ D4-branes in Figure 8A cannot move any
directions,
the tilted $(N_c'-N_c''-\widetilde{N}_c)$-flavor D4-branes can move 
$w$ direction in Figure 8B(and its mirrors).
The remaining upper $\widetilde{N}_c$ D4-branes are fixed also and cannot 
move any direction. 
Note that 
there is a decomposition 
\bea
N_c'-N_c''=(N_c-N_c'-N_c''-4)+\widetilde{N}_c.
\nonu
\eea

The brane configuration for zero mass for the bifundamental,
which has only a cubic superpotential,
can be obtained from Figure 8A by moving
the upper  NS5'-brane together with $(N_c'-N_c'')$ color D4-branes 
into the origin $v=0$(and their mirrors).
Then the number of dual colors for D4-branes 
becomes $\widetilde{N}_c$ between the $NS5_L$-brane and its mirror, 
 $N_c'$ between $NS5_L$-brane and the NS5'-brane
and $N_c''$ 
between NS5'-brane and $NS5_R$-brane.

The brane configuration in Figure 8A is stable as long as the
distance $\Delta x$ between the upper $NS5_L'$-brane and 
the middle NS5'-brane is large. If they are close to 
each other then this brane
configuration is unstable to decay and it becomes 
the meta-stable brane configuration in Figure
8B.
Since the two NS5'-branes are located at different sides of NS5-brane
in Figure 8B, contrary to the previous cases,
for the DBI computation, this fact should be taken into account. 
One can regard these brane configurations as particular states in the
magnetic gauge theory with the gauge group (\ref{dualdual}) and
superpotential (\ref{super}).
When the two NS5'-branes which are connected by $\widetilde{N}_c$
D4-branes 
are replaced by two coincident $(N_c'-N_c'')$
D6-branes,
the brane configuration of Figure 8B(with the rotation of 
$NS5_R$-brane by $\frac{\pi}{2}$) is the same as the Figure 7(with
opposite movement of D6-branes to $v$ direction) studied in 
\cite{Ahn07-4}
 where the gauge group is   
given by 
$SU(2n_f+2n_c'-n_c+4) \times SU(n_c')$ 
with $n_f$ fundamentals, bifundamentals, an antisymmetric flavor, a
conjugate symmetric flavor, and various gauge singlets.
Then the our $N_c$ corresponds to the $n_c$,
the number $(N_c'-N_c'')$ corresponds to $n_f$,
and 
our $N_c''$ corresponds to the $n_c'$ of \cite{Ahn07-4}.

The $N_c''$ D4-branes between the $NS5_L$-brane and the $NS5_R$-brane can
slide along the $v$ direction(and their mirrors to the opposite direction).

$\bullet$ Gauge theory at small $\Delta x$ 

The gauge couplings for the  gauge group factors are
given by similarly
and the superpotential 
corresponding to Figures 8A and 8B is given by
\bea
W_{dual} = h \Phi' f \widetilde{f} - h \mu^2 \tr \Phi' + \hat{q}
 \widetilde{s} \hat{q} , \qquad
h^2=  g_{2,mag}^2
\nonu
\eea
and the  mass parameter $\mu^2$ is the same as before.
See also the relevant papers \cite{Ahn07-1,Ahn07-5}.
Then the product $ f \widetilde{f}$ is 
a $\widetilde{N}_c \times \widetilde{N}_c$ 
matrix where the second gauge group indices for $f$ and $\widetilde{f}$ 
are contracted with those
of $\Phi'$ while $\Phi'$ is a 
$(N_c'-N_c'') \times (N_c'-N_c'')$ matrix.
Although the field $f$ itself is an antifundamental in the second gauge
group,
the product $f \widetilde{f}$ has the same representation with the 
product of dual quarks
and the second gauge group indices for the field $\Phi'$ play the
role of the flavor indices.

Therefore, the F-term equation, the derivative of $W_{dual}$ with respect to the
meson field $\Phi'$ cannot be satisfied if the $(N_c'-N_c'')$ exceeds
$\widetilde{N}_c$.
So the supersymmetry is broken.   
The classical moduli space of vacua can be obtained from F-term
equations. 
That is, 
there are five equations from F-term conditions:
$
f^a \widetilde{f}_b -\mu^2 \delta^a_b =0,  \Phi' f =0,  \widetilde{f}
\Phi'=0, \hat{q} \widetilde{s} =0$, and $ \hat{q} \hat{q} =0$.
Then the solutions for these
are given by 
\bea
<f>   & = & 
\left(
\begin{array}{c}
\mu   {\bf 1}_{\widetilde{N}_c}  \nonu \\
0
\end{array}
\right), \quad
<\widetilde{f}>   = 
\left(
\begin{array}{cc}
\mu   {\bf 1}_{\widetilde{N}_c} & 0 \nonu \\
\end{array}
\right), 
\quad
<\Phi'> =
 \left(
\begin{array}{cc}
0  & 0
 \\
0 & \Phi_0'  {\bf 1}_{(N_c'-N_c''-\widetilde{N}_c)} 
\end{array}
\right),
\nonu \\
<\hat{q}> & = & 0,   \quad 
<\widetilde{s}> = 0.
\nonu
\eea
One can expand around the solutions, as done in \cite{Ahn07-5}. Although there
exists an extra last term  in (\ref{super}), this does not contribute to the
one loop result.
At one loop, the effective potential $V_{eff}^{(1)}$ for $\Phi_0'$
leads to the positive value for $m_{\Phi_0'}^2$ implying that these
vacua are stable.
The gauge theory analysis 
is only valid in the regime where 
$\Delta x$ is smaller than $\exp(-\frac{C}{g_s})$ with some positive
constant $C$. 

\subsection{
${\cal N}=1$ 
$SU(N_c) \times SU(\widetilde{N}_c') \times SU(N_c'')$ magnetic theory
}

Let us consider the Seiberg dual for the second gauge group and take 
the mass deformation by moving the $N_c''$ D4-branes betwee the
the NS5-brane and $NS5_R'$-brane to $+v$ direction.
See also the Figure 9. Since the discusssion is analogous to previous
subsection 2.1, we mention the main results briefly. 

$\bullet$ Seiberg dual and mass deformation

By applying the Seiberg dual to the second gauge group $SU(N_c')$ factor in 
(\ref{gag1})  and 
interchanging the $NS5_L$-brane and the NS5'-brane(and their mirrors)
from Figure 7A,
one obtains the Figure 9A.

\begin{figure}[ht]
   \epsfxsize=5.0in 
\centerline{\epsffile{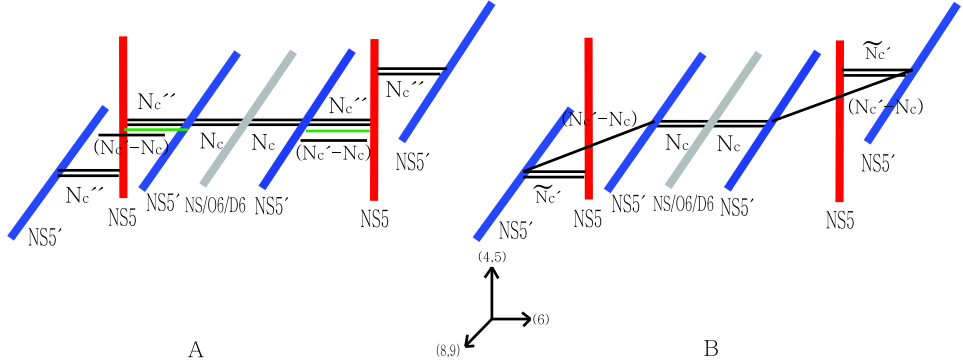}}
   \caption[FIG. \arabic{figure}.]{ 
The  ${\cal N}=1$ 
magnetic brane configuration for the gauge group $SU(N_c) \times 
SU(\widetilde{N}_c'=N_c+N_c''-N_c') \times
SU(N_c'')$ 
with D4-
and $\overline{D4}$-branes(9A) and with 
a misalignment between D4-branes(9B) when the NS5'-branes are close to
each other.
 The number of tilted D4-branes is equal to $N_c'-N_c=
N_c''-\widetilde{N}_c'$ in 9B.
 }
\end{figure}

When two NS5'-branes in Figure 9A are close to each other, it becomes 
the meta-stable brane configuration Figure 9B
 by realizing that the number of $N_c''$
D4-branes connecting between NS5-brane and $NS5_R'$-brane can
be rewritten as $(N_c'-N_c)$ plus $\widetilde{N}_c'$. 
Now 
we draw this ${\cal N}=1$ supersymmetric magnetic
brane configuration in Figure 9.
We assume that the number of colors satisfies (\ref{color}).
The brane configuration consisting of NS5-brane and two NS5'-branes in
Figure 9 is
exactly the same as those in Figure 15 of \cite{Ahn07-6} if we
replace the NS5/O6/D6-branes with $O6^{+}$-plane. 
There is an other mass deformation 
by  moving the middle NS5'-brane located at $x^6=0$ to $+v$ direction. 

The dual gauge group is given by (\ref{dualnew}) and for convenience
we rewrite it here
\bea
SU(N_c) \times SU(\widetilde{N}_c'=N_c+N_c''-N_c') \times SU(N_c''). 
\label{dualnew11}
\eea
The matter contents are the field $f$ 
 charged under
$({\bf N_c, \overline{\widetilde{N}_c'}, 1})$, a field $g$ charged under
$({\bf 1, \widetilde{N}_c', \overline{N_c''}})$, and their conjugates 
$\widetilde{f}$ and $\widetilde{g}$ under the dual gauge group
(\ref{dualnew11})
and  
the gauge-singlet $\Phi''$ for the second dual gauge group in the 
adjoint representation for the third dual gauge group, 
i.e.,  ${(\bf 1, 1, {N_c''}^2-1)  \oplus (1,1,1)}$ under the 
dual gauge group.
Then the  $\Phi''$ is a $N_c'' \times N_c''$ matrix.
There are also
the antisymmetric flavor $A$, the conjugate 
symmetric flavor $\widetilde{S}$ and eight fundamentals $\hat{Q}$ for 
$SU(N_c)$.

$\bullet$ Meta-stable brane configuration


The brane configuration for zero mass for the bifundamental,
which has only a cubic superpotential,
can be obtained from Figure 9A by moving
the upper  NS5'-brane together with $N_c''$ color D4-branes 
into the origin $v=0$(and their mirrors).
Then the number of dual colors for D4-branes 
becomes $N_c$ between the $NS5_L'$-brane and its mirror, 
 $\widetilde{N}_c'$ between $NS5_L'$-brane and NS5-brane
and $N_c''$ 
between NS5-brane and $NS5_R'$-brane.

The brane configuration in Figure 9A is stable as long as the
distance $\Delta x$ between the upper NS5'-brane and 
the lower NS5'-brane 
is large. If they are close to each other, then this brane
configuration is unstable to decay to 
the brane configuration in Figure
9B.
One can regard these brane configurations as particular states in the
magnetic gauge theory with the gauge group and
superpotential.
The  upper tilted $(N_c''-\widetilde{N}_c')$ flavor D4-branes of 
straight brane configuration
of
Figure 9B  bend since there exists an attractive
gravitational interaction
between those flavor D4-branes and NS5-brane from the DBI action. 
As mentioned in \cite{Ahn07-5},
the two NS5'-branes are located at different side of NS5-brane in
Figure 9B and the DBI action computation for this bending curve
should be taken into account. 

The $N_c$ D4-branes between the $NS5'_L$-brane and its mirror can
slide along the $w$ direction.
When $y_1$ goes to zero, this meta-stable brane configuration of
Figure 9B where there are three NS5'-branes at the origin 
reduces to  the Figure 6 of \cite{Ahn07-5} where there is a single
NS5'-brane at the origin.

$\bullet$ Gauge theory at small $\Delta x$ 

The low energy dynamics of the magnetic brane configuration 
can be described by the ${\cal N}=1$ supersymmetric gauge theory
with gauge group (\ref{dualnew11})
and the gauge couplings for the three gauge group factors are
given by the expressions in subsection 2.1.

The dual gauge theory has  a meson field  $\Phi''$  and 
bifundamentals $f, \widetilde{f}, g$ and $\widetilde{g}$ under the dual gauge
group (\ref{dualnew11}) and the superpotential 
corresponding to Figures 9A and 9B is given by 
the expressions in subsection 2.1.
Then $ g \widetilde{g}$ is a $\widetilde{N}_c' \times \widetilde{N}_c'$ 
matrix where the third gauge group indices for $g$ and $\widetilde{g}$ 
are contracted with those
of $\Phi''$ while $\Phi''$ is a 
$N_c'' \times N_c''$ matrix.
The product $g \widetilde{g}$ has the same representation for the 
product of quarks
and moreover, 
the third gauge group indices for the field $\Phi''$ play the
role of the flavor indices.

When the upper NS5'-brane(or $NS5_R'$-brane) 
is replaced by coincident $N_c''$ 
D6-branes in Figure 9B, this brane configuration looks similar to the
Figure 7 
found in \cite{Ahn07-4} where the gauge group was given by 
$SU(n_c) \times SU(n_f'+n_c-n_c')$ 
with $n_f'$ fundamentals, bifundamentals, an antisymmetric flavor, a
conjugate symmetric flavor, and gauge singlets. 
Then the present $N_c$ corresponds to the $n_c$,
$N_c'$ corresponds to $n_c'$,
and 
$N_c''$ corresponds to the $n_f'$ of \cite{Ahn07-4}. 
The location of $NS5_R'$-brane is in the right hand side of NS5-brane
while the location of $n_f'$ D6-branes of Figure 7 in \cite{Ahn07-4}
is in the left hand side of NS5-brane.


\subsection{${\cal N}=1$ 
$SU(N_c) \times SU(N_c') \times SU(\widetilde{N}_c'')$ magnetic theory}

Let us consider the Seiberg dual for the third gauge group and take 
the mass deformation by moving the $(N_c'-N_c)$ D4-branes between 
the $NS5_L'$-brane and the NS5-brane
to $+v$ direction. See also the Figure 10.
 Since the discusssion is analogous to previous
subsection 2.2, we mention the main results briefly.

$\bullet$ Seiberg dual and mass deformation by $G$ and $\widetilde{G}$

By applying the Seiberg dual to the third gauge group $SU(N_c'')$ factor in 
(\ref{gag1})  and 
interchanging the NS5'-brane and the $NS5_R$-brane(and their mirrors),
one obtains the Figure 10A.


When two NS5'-branes in Figure 10A are close to each other, then 
it leads to Figure 10B
 by realizing that the number of $(N_c'-N_c)$
D4-branes connecting between $NS5_L'$-brane and NS5-brane can
be rewritten as $(N_c''-N_c)$ plus $\widetilde{N}_c''$.
The brane configuration consisting of NS5-brane and two NS5'-branes in
Figure 10B is
exactly the same as the last three NS-branes 
in Figure 5B'' of \cite{Ahn07-6}. 
The brane configuration in
Figure 10 is
exactly the same as those in Figure 16 of \cite{Ahn07-6} if we
replace the NS5/O6/D6 with $O6^{+}$-plane. 

\begin{figure}[ht]
   \epsfxsize=5.0in 
\centerline{\epsffile{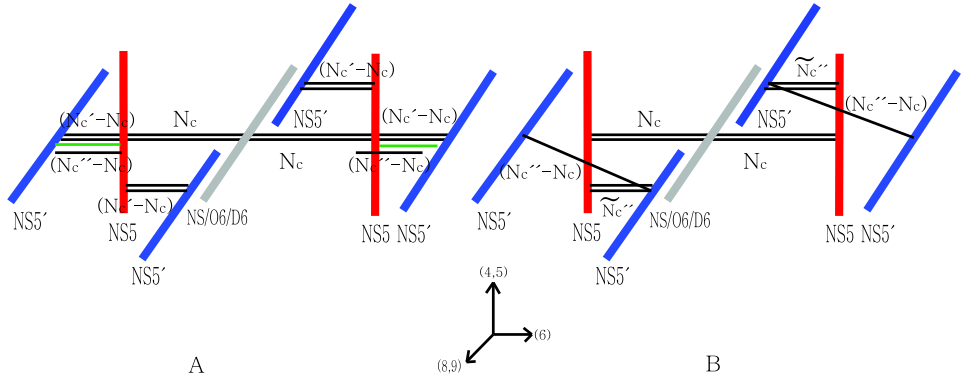}}
   \caption[FIG. \arabic{figure}.]{ 
The  ${\cal N}=1$ 
magnetic brane configuration for the gauge group $SU(N_c) \times 
SU(N_c') \times
SU(\widetilde{N}_c''=N_c'-N_c'')$ 
with D4-
and $\overline{D4}$-branes(10A) and with 
a misalignment between D4-branes(10B) when the NS5'-branes are close to
each other.
 The number of tilted D4-branes is equal to $N_c''-N_c=
(N_c'-N_c)-\widetilde{N}_c''$ in 10B.  }
\end{figure}

The dual gauge group is given by (\ref{dualneww}) and we repeat here
\bea
SU(N_c) \times SU(N_c') \times SU(\widetilde{N}_c''=N_c'-N_c''). 
\label{dualneww11}
\eea
The matter contents are the field $F$ 
 charged under
$({\bf N_c, \overline{N_c'}, 1})$, a field $g$ charged under
$({\bf 1, N_c', \overline{\widetilde{N}_c''}})$ 
 and their conjugates 
$\widetilde{F}$ and $\widetilde{g}$
under the dual gauge group
(\ref{dualneww11})
and  
the gauge-singlet $\Phi'$ which is in the 
adjoint representation for the second dual gauge group, 
in other words,   
$ ({ \bf   1,  (N_c'-N_c)^2-1,1})  \oplus  ({\bf 1,1,1})$ under the 
dual gauge group (\ref{dualneww11})  where the gauge group $SU(N_c')$
is broken 
to $SU(N_c'-N_c)$.
Then the $\Phi'$ is a $(N_c'-N_c) \times (N_c'-N_c)$ matrix.
Only $(N_c'-N_c)$ D4-branes are participating in the mass deformation.
There are also
the antisymmetric flavor $A$, the conjugate 
symmetric flavor $\widetilde{S}$ and eight fundamentals $\hat{Q}$ for 
$SU(N_c)$.

$\bullet$ Meta-stable brane configuration


The brane configuration for zero mass for the bifundamental,
which has only a cubic superpotential,
can be obtained from Figure 10A by moving
the upper  NS5'-brane together with $(N_c'-N_c)$ color D4-branes 
into the origin $v=0$(and their mirrors).
Then the number of dual colors for D4-branes 
becomes $N_c$ between the $NS5_L'$-brane and its mirror, 
$N_c'$ between the $NS5_L'$-brane and the  NS5-brane
and $\widetilde{N}_c''$ 
between NS5-brane and $NS5_R'$-brane.

When the upper NS5'-brane(or $NS5_L'$-brane) 
is replaced by coincident $(N_c'-N_c)$ 
D6-branes in Figure 10B, this brane configuration looks similar to the
Figure 5
found in \cite{Ahn07-4} where the gauge group was given by 
$SU(n_c) \times SU(n_f'+n_c-n_c')$ 
with $n_f'$ fundamentals, bifundamentals, a symmetric flavor, a conjugate
symmetric flavor, and gauge singlets. 
The only difference between these two brane configurations 
appears at the origin $x^6=0$.
Then the present $N_c$ corresponds to the $n_c$,
$(N_c'-N_c)$ corresponds to $n_f'$,
and 
$N_c''$ corresponds to the $n_c'$ of \cite{Ahn07-4}. 


$\bullet$ Gauge theory at small $\Delta x$ 

The low energy dynamics of the magnetic brane configuration 
can be described by the ${\cal N}=1$ supersymmetric gauge theory
with gauge group (\ref{dualneww11})
and the gauge couplings for the three gauge group factors are
given by
the expressions in subsection 2.2.
The dual gauge theory has  a meson field  $\Phi'$  and 
bifundamentals $f, \widetilde{f}, g$ and $\widetilde{g}$ under the dual gauge
group (\ref{dualneww11}) and the superpotential 
corresponding to Figures 10A and 10B is given by 
the one in subsection 2.2.
Then $ g \widetilde{g}$ is a $\widetilde{N}_c'' \times \widetilde{N}_c''$ 
matrix where the second gauge group indices for $g$ and $\widetilde{g}$ 
are contracted with those
of $\Phi'$ while $\Phi'$ is a 
$(N_c'-N_c) \times (N_c'-N_c)$ matrix.
The product $g \widetilde{g}$ has the same representation for the 
product of quarks
and moreover, 
the second gauge group indices for the field $\Phi'$ play the
role of the flavor indices.


\section{Conclusions and outlook}

The meta-stable brane configurations we have found are summarized by
Figures 3, 5, 6, 8, 9, 10.
If we replace the upper NS5'-brane in Figures 6B(with a rotation of
$NS5_R'$-brane) and 8B(with a rotation of $NS5_R$-brane) 
with the coincident D6-branes, 
those brane configurations become nonsupersymmetic
minimal energy brane configurations in 
the Figures 4 and 6 found in \cite{Ahn07-4} previously.
By changing some of the NS-branes, D-branes or O-planes 
from the remaining 
Figures 3, 5, 9 and 10, one obtains the brane configurations Figures 12,
14, 15 and 16 given by 
\cite{Ahn07-6} respectively.

Some different directions on the meta-stable vacua
are present in
recent relevant works \cite{HV}-\cite{FU} where 
some of them are described in the type IIB string theory.
It would be very interesting to find out
how the meta-stable brane configurations from 
type IIA string theory including the present work are related to
those brane configurations from type IIB string theory.

\vspace{.7cm}

\centerline{\bf Acknowledgments}

I would like to thank KIAS(Korea Institute for Advanced Study)
and Harvard High Energy Theory Group 
for hospitality where this work was undertaken.
This work was supported by grant No.
R01-2006-000-10965-0 from the Basic Research Program of the Korea
Science \& Engineering Foundation.

\end{document}